\documentclass[amsmath,
amssymb,
nofootinbib,
notitlepage,
floatfix,
prd,
superscriptaddress,
twocolumn,
longbibliography,
]{revtex4-1}
\usepackage[utf8]{inputenc}
\usepackage[T1]{fontenc}
\usepackage[usenames,dvipsnames]{xcolor}
\usepackage{graphicx,hyperref,orcidlink,float}
\hypersetup{colorlinks=true,citecolor=RoyalBlue,linkcolor=RoyalBlue,urlcolor=RoyalBlue}
\usepackage{lipsum}
\usepackage{comment}
\usepackage{appendix}
\usepackage{graphicx}
\usepackage{soul}
\usepackage{tikz}
\usepackage{pgfplots}
\usepackage{dcolumn}
\usepackage{bm}
\usepackage{amsmath}
\usepackage{listings}
\usepackage[capitalize]{cleveref}
\usepackage{slashed}
\usepackage{enumitem}
\usepackage{subcaption}
\usepackage{ragged2e} 

\newcommand{\msun}{M_{\odot}}

\newcommand{\BF}[1]{\boldsymbol{#1}}

\newcommand{\bftr}[1]{\boldsymbol{#1}_{\mathrm{tr}}}

\newcommand{\bfml}[1]{\boldsymbol{#1}_{\mathrm{ML}}}

\graphicspath{{figures/}}
\begin{document}

\title{Millilensing induced systematic biases in parameterized tests of General Relativity} %

  \author{Anna Liu}%
\email{ania.liu@link.cuhk.edu.hk}%
  \affiliation{Department of Physics, The Chinese University of Hong Kong, Shatin, NT, Hong Kong}%
\author{Rohit S. Chandramouli}%
\email{rsc4@illinois.edu}%
  \affiliation{Illinois Center for Advanced Studies of the Universe \& Department of Physics\\
   University of Illinois at Urbana-Champaign, Urbana, Illinois 61801, USA}%
   \affiliation{SISSA, Via Bonomea 265, 34136 Trieste, Italy and INFN Sezione di Trieste}%
  \author{Otto A. Hannuksela}%
  \affiliation{Department of Physics, The Chinese University of Hong Kong, Shatin, NT, Hong Kong}%
  \author{Nicol\'as Yunes}%
  \affiliation{Illinois Center for Advanced Studies of the Universe \& Department of Physics\\
   University of Illinois at Urbana-Champaign, Urbana, Illinois 61801, USA}%
  \author{Tjonnie G. F. Li}%
  \affiliation{Department of Physics, The Chinese University of Hong Kong, Shatin, NT, Hong Kong}%
  \affiliation{Institute for Theoretical Physics, KU Leuven, Celestĳnenlaan 200D, B-3001 Leuven, Belgium }%
  \affiliation{Department of Electrical Engineering (ESAT), KU Leuven, Kasteelpark Arenberg 10, B-3001 Leuven, Belgium}

\date{\today}

\begin{abstract}
\noindent
    Tests of general relativity (GR) can be systematically biased when our waveform models are inaccurate.
    We here study systematic biases in tests of general relativity induced by neglecting lensing effects for millilensed gravitational-wave signals, where the lens mass is typically in the $10^3\msun$--$10^5\msun$ range.
    In particular, we use a nested-sampling Bayesian parameter estimation and model selection analysis of a millilensed signal with an unlensed parameterized post-Einsteinian (ppE) recovery model. 
    We find that the ppE model is significantly biased toward a detection of a deviation from general relativity at signal-to-noise ratios of 30 and higher, especially when the source is aligned with the lens mass (the lensing effect is pronounced) and when its total mass is low (the signal duration is long).
    We use a toy model and the linear signal and Laplace approximations to provide a semi-analytic explanation for the trends in the systematic errors found in the nested sampling analysis.
    Moreover, a Bayes factor analysis reveals that the (unlensed) ppE model is weakly favored over the (unlensed) GR model, and a fitting factor study shows there is a significant loss of signal-to-noise ratio when using the (unlensed) ppE model. 
    This implies that although a parameter estimation study may incorrectly infer a deviation from general relativity, a residual signal-to-noise ratio test would reveal that the ppE model is not a good fit to the data.
    Thus, with current detectors, millilensing-induced systematic biases are unlikely to result in false positive detections of GR deviations.
\end{abstract}
\maketitle

\section{\label{sec:introduction}Introduction}
\noindent
Gravitational-wave (GW) observations have emerged as a powerful tool for probing fundamental physics, astrophysics, cosmology, and nuclear physics.
A treasure trove of new GW data is being collected, with 90 confirmed detections as of the third observing run, and about another 100 in the first half of the ongoing fourth observing run O4~\cite{abbott2016improved, abbott2016observation, abbott2019gwtc1, abbott2021gwtc2, abbott2021gwtc3}.
The large majority of GW observations are from the mergers of quasi-circular binary black holes (BBH), and a few come from binary neutron star (BNS) or neutron star/black hole binary (NSBH) systems.
We have yet to convincingly and unambiguously detect GWs from more complex coalescence events, such as eccentric and highly-precessing mergers of binaries with asymmetric masses.
If such waves were present in the data, they would pose challenges for accurate parameter estimation (PE), as there is a lack of accurate, precise, and complete (inspiral-merger-ringdown) waveform models in certain regions of parameter space. 
Imperfect waveform models due to mismodeled or unmodeled physical effects can lead to systematic parameter estimation biases or even force us to miss GWs altogether. 
In particular, systematic mismodeling biases can affect parameterized tests of general relativity (GR), potentially leading to false positive detections of GR deviations~\cite{Gupta:2024gun,rohit:syspaper1}.

In spite of these potential problems, all observed GW observations have been used to test GR in the strong-field, dynamical regime, achieving one of the primary goals of GW analysis.
No deviations from GR have been found to date~\cite{abbott2019tests, abbott2021TGR, abbott2021tests}, and thus, constraints on the magnitude of such deviations have been placed, allowing various modified theories of gravity to be tested, see e.g.~\cite{Belgacem:2017ihm, Perkins:2021mhb, Chua:2020oxn, Carullo:2021dui,  Gupta:2020lxa, Hagihara:2019ihn,  Johnson-McDaniel:2021yge, Kramer:2021jcw, Nishizawa:2017nef, Shoom:2021mdj,  Wang:2021yll, Wong:2021cmp, Yunes:2016jcc, Nair:2019iur, Silva:2020acr, Perkins:2021mhb, Schumacher:2023cxh, Callister:2023tws, Lagos:2024boe, Liu:2024atc, Yunes:2024lzm}.
A successful way of testing GR is via parameterized tests, wherein a deviation or a set of deviations from GR are introduced into the waveform model, as for example done in the parameterized post-Einsteinian (ppE) framework~\cite{Yunes_2009} (see~\cite{Chatziioannou:2012rf,Moore:2020rva,Perkins:2022fhr,Mehta_2023,Loutrel:2022xok,Mezzasoma:2022pjb,Xie:2024ubm} for various ppE extensions and~\cite{Li_2012,agathos2014tiger,Perkins:2022fhr} for various data analysis implementations of the ppE framework).
By doing so, a constraint on the deviation parameter allows for a constraint on families of modified gravity theories and associated phenomena~\cite{Yunes_2016}. 
Such tests of GR are robust only when the GR waveform model is sufficiently accurate for describing the signal. 
Understanding how tests of GR can be biased due to different sources of systematic error is one of the latest challenges in GW inference~\cite{Gupta:2024gun}.

Many sources of waveform mismodeling errors affect tests of GR (see for e.g.~\cite{Gupta:2024gun,rohit:syspaper1}).
Recently, there has been interest in studying the biases resulting from neglecting the effects of gravitational lensing. 
Similar to the electromagnetic-wave case, GWs are subject to gravitational lensing when they pass near massive objects~\cite{ohanian1974focusing, Deguchi1986WaveRadiation, Wang:1996, Nakamura:1997sw,Takahashi2003WaveBinaries}. 
While anticipated to become detectable based on theoretical models and simulations as the sensitivity of detectors is improved~\cite{Ng2017PreciseHoles,Oguri2018EffectMergers,Li2018GravitationalPerspective,Xu2021PleasePopulations, Wierda2021BeyondLensing,More2021ImprovedEvents,Smith2022DiscoveringObservatory,Phurailatpam:2024enk}, lensed GWs have not yet been confidently observed~\cite{Hannuksela2019SearchEvents, Dai2020SearchO2, abbott2021search, Basak2022ConstraintsMicrolensing, Kim2022Deep-2, abbott2023search, Janquart_2023, Goyal:2023lqf, Janquart:2024ztv}. 
However, if a signal were lensed, it could lead to several observational signatures, such as magnification~\cite{Dai2016EffectMergers,Broadhurst2018ReinterpretingDistances,Hannuksela2019SearchEvents,Pang2020LensedDetection, Shan:2020esq}, multiple images~\cite{Haris2018IdentifyingMergers}, complex phase shifts~\cite{Dai2017OnWaves, ezquiaga2021phase}, frequency-dependent wave optics modulations~\cite{Takahashi2003WaveBinaries, cao2014gravitational, Lai2018DiscoveringLensing, Christian:3Gdetections, Jung:2019compactDM, diego2019observational, Mishra2021GravitationalGalaxies, meena2020gravitational, Cheung2020Stellar-massWaves, Bulashenko2021LensingPatternb, cremonese2021breaking, Seo2021StrongMicrolensing, Yeung2021MicrolensingMacroimages, Qiu2022AmplitudeWaves, Wright2021Gravelamps:Selection, Meena2022GravitationalPopulation,Shan:2023ngi}, and beating patterns due to millilensing (waveform superposition in the Eikonal limit)~\cite{Liu2023Exploring,Leung:2023lmq}. 
Neglecting lensing effects on the gravitational waveform could lead to systematic biases, as shown recently in~\cite{mishra2023unveiling, wright2024effect}.
In particular, this kind of systematic bias could influence tests of GR, deteriorating them or even leading to the (false-positive) detection of GR deviations (see e.g.~\cite{ezquiaga2020gravitational, Ezquiaga2022ModifiedLensing, Li_2023, gupta2024possible}). 
Studying how tests of GR could be biased due to the neglect of lensing effects in the recovery model is therefore important when characterizing and quantifying systematic biases in parameter estimation.

The lensing of GWs can be broadly categorized into \textit{strong lensing} and \textit{micro lensing}, depending on the lens mass and its impact on the frequency evolution of the GW signals.
We do not consider weak lensing (which changes the overall GW amplitude), in this context, which would be induced by fluctuations in the background matter density and is expected to mostly affect signals from high-redshift sources only ($z\gtrsim 1$)~\cite{Holz_2005, Wu_2023}, thus becoming relevant for future, space-based detectors, such as LISA~\cite{amaro2017laser}, TianQin~\cite{Luo_2016} or Taiji~\cite{TaijiScientific:2021qgx}.
Lenses with masses of $\mathcal{O}(10^6 M_\odot)$ and higher, such as galaxies and galaxy clusters, lead to strong gravitational lensing, which results in multiple lensed GW signals separated by time delays ranging from minutes to years for transient GWs detectable by ground-based detectors ~\citep{Smith2017WhatClusters, Smith2018DeepGW170814, Haris2018IdentifyingMergers, Liu2020IdentifyingVirgo, Robertson:2020, Ryczanowski:2020, Dai2020SearchO2, Wang2021IdentifyingDetectors, ezquiaga2021phase, Lo2021ASignals, Janquart2021AEvents, Janquart2021OnModes, Vijaykumar2022DetectionSignals, ccalicskan2022lensing, Cao2022DirectWaves, Cheung:2023uzy, Poon:2024zxn, Uronen:2024bth, Leong:2024nnx}.
Smaller-mass lenses, such as stars or stellar-mass compact objects, can cause micro- and millilensing, with multiple lensed signals \textit{overlapping} with each other, resulting in beating patterns in the frequency evolution of the GW~\citep{Takahashi2003WaveBinaries, cao2014gravitational, Lai2018DiscoveringLensing, Christian:3Gdetections, Jung:2019compactDM, diego2019observational, Mishra2021GravitationalGalaxies, meena2020gravitational, Cheung2020Stellar-massWaves, Bulashenko2021LensingPatternb, cremonese2021breaking, Seo2021StrongMicrolensing, Yeung2021MicrolensingMacroimages, Qiu2022AmplitudeWaves, Wright2021Gravelamps:Selection, Meena2022GravitationalPopulation, Shan:2023ngi, Shan:2023qvd, Shan:2024min}.
For microlensed GW signals in the wave optics regime (i.e.~when the characteristic GW wavelengths are longer than the Schwarzschild radius of the lens, $\lambda_{\rm{GW}}\gtrsim R_{\rm{Sch}}$), the lens masses are typically below $10^3 \msun$.
Meanwhile, lens masses in the range $10^3 M_\odot-10^5 M_\odot$ result in millilensed GW signals, described by the geometric optics regime (i.e.~when $\lambda_{\rm{GW}}\lesssim R_{\rm{Sch}}$).

In this paper, we focus on systematic biases arising if we were to detect a millilensed GW signal (as described in GR) without knowing it, and we were to analyze it with a recovery model that neglects lensing effects but allows deviations from GR.  
In particular, we focus on addressing the following questions:
\begin{enumerate}
    \setlength\itemsep{0.5em}
    \item For which signal and lens parameters are inspiral ppE tests of GR biased when lensing effects are neglected from the recovery model?
    \item Which ppE tests are more significantly biased than others, and what is the characterization of the bias?
\end{enumerate}
Answering these questions helps us determine what kind of GR deviation we would mistakenly think we have detected,
when the signal is actually consistent with GR but it is millilensed. 
Moreover, the answers to these questions also indicate which systems and lenses require our utmost caution, as they are more likely to be confused with GR deviations.  

To address these questions, we perform Bayesian inference and model selection studies on a varied set of injected millilensed GW signals built within GR and recovered with a ppE, unlensed waveform model. 
Both the injections and recovery models are constructed from an inspiral-merger-ringdown (IMR) PhenomPv2 model~\cite{Hannam:2013oca,Schmidt:2012rh}, which is then either lensed (for the injections) or modified with ppE terms (for the recovery). 
The PhenomPv2 model is an accurate description of the GWs emitted in the spin-precessing, quasi-circular inspiral, merger and ringdown of BBHs. By varying the injected total mass, signal-to-noise ratio (SNR), phenomenological lens parameters, and the post-Newtonian (PN)\footnote{The PN approximation is one in which the field equations are expanded and solved in powers of $v/c$, where $v$ is the orbital velocity of the binary and $c$ is the speed of light~\cite{Blanchet:2013haa}. A term of $N$PN order scales as $(v/c)^{2 N}$ relative to the leading-order term in the PN series.} 
order of the ppE deviation in the recovery model, our results offer a robust understanding of how neglecting lensing effects impact the biases in tests of GR. 
Furthermore, we develop a simple toy model (similar to~\cite{rohit:syspaper1}) to better understand the properties of the systematic biases due to neglecting millilensing. 
Using the linear signal approximation (LSA)~\cite{Cutler_2007,Flanagan:1997kp} and the Laplace approximation~\cite{gregory_2005}, we obtain analytic expressions for the systematic bias in the ppE parameter, the fitting factor of the ppE model, and the Bayes factor between the ppE and GR models.
By doing so, we illustrate how the biases depend on the PN order of the ppE deviation, as well as the lensing strength, and recover the same trends shown by the results of our numerical nested sampling studies.

Our study establishes the following facts. 
As one may have expected, the systematic bias in tests of GR becomes problematic when 
(i) the lensing distortions are significant, 
(ii) the signal is loud (high SNR), and 
(iii) the total mass $m$ of the GW source is low (long signal duration). 
Importantly, systematic errors are largely independent of SNR, while statistical errors decrease inversely with SNR, leading to a growing ratio of systematic to statistical error as SNR increases.
We find that already at SNRs of 30, the systematic error can exceed the statistical error, provided the lensing distortions of the signal are significant (with a source offset position relative to the lens $\xi$, divided by the dimensional Einstein radius of the lens $r_E$, of $y=\xi/r_E=0.3$) 
and the binary mass is around $20 M_\odot$. 
For a fixed signal (source and lensing) configuration, the absolute systematic error typically increases with the PN order of the ppE deviation. 
Similarly, for a given PN order of the ppE deviation and lensing configuration, the absolute systematic error increases with the total mass of the source. 
However, there is a lack of a clear trend in the significance of the bias (ratio of systematic to statistical error) as the statistical error also depends on the total mass, PN order of the ppE deviation, and lensing strength.
For instance, when the lensing distortions are significant and when testing for GR deviations at -1PN order, the small mass ($20 \msun$) injection exhibits more significant bias than the corresponding large mass ($40 \msun$) case, as expected.
However, when recovering with a 1PN ppE deviation given significant lensing distortions, the large mass ($40 M_{\odot}$) system exhibits a more significant bias than the small binary mass system.
Therefore, the main (conservative) conclusions of our analysis (also summarized in Sec.~\ref{subsec:executive_summary}) are that utmost care must be taken when testing for ppE deviations at low PN orders with low total-mass ($m \simeq 20 M_{\odot}$) signals, and when testing for higher PN order deviations with high total-mass ($ m \simeq 40 M_{\odot}$) signals, and with high SNRs ($\geq 30$).

When the lensing distortions are significant ($y=0.3$), we find that the absolute ratio of systematic to statistical error in the ppE deviation is larger than 1 at -1PN, 1PN, and 2PN orders, implying significant systematic biases in the ppE tests. 
However, even in this case when there is significant lensing distortions and where Bayesian parameter estimation leads to an incorrect identification of a GR deviation, we also find that the fitting factor between the signal and the maximum-likelihood ppE model is about $0.81$. 
Consequently, a residual SNR test (where one calculates the SNR of the difference between the signal and the maximum likelihood model) would fail, revealing that a non-GR model is not a good fit to the data, and thus, the data is not necessarily consistent with a GR deviation. 
Furthermore, the Bayes factor at an SNR of 30 does not strongly favor the ppE model over the GR model (a model selection test), but it does so at SNR of 60, implying that a model selection test will pass at the larger SNR.
Since the fitting factor does not scale with SNR, even if the model selection test is passed, the residual SNR test will fail, revealing that the detected deviation from GR stems from waveform mismodeling.
Therefore, this concrete example of millilensing considered in our work emphasizes the importance of always concurrently running residual tests of GR together with parametric tests of GR (including model selection).
Our work also highlights the necessity of accounting for different astrophysical effects, such as the GW lensing presented here, when testing GR.

Ours is not the first study of systematic bias due to the neglect of lensing features in the waveform when testing GR. 
Reference~\cite{mishra2023unveiling} demonstrated that wave optics effects in microlensed GWs can lead to biases in both parameterized tests of GR and IMR consistency tests, when such microlensing effects are neglected from the recovery waveform model. 
On the flip side, Ref.~\cite{wright2024effect} established that GWs carrying imprints of deviations from GR can be misinterpreted as lensed GW signals.
Our study primarily differs from~\cite{mishra2023unveiling} by studying biases incurred from analysis of phenomenological millilensed GW signals and by investigating biases in inspiral tests of GR at -1PN, 1PN, and 2PN orders.
Furthermore, our work incorporates the decision scheme outlined in~\cite{rohit:syspaper1}, which uses the Bayes factor between the non-lensed ppE and GR recoveries, as well as the fitting factor of the non-lensed ppE recovery, to assess the significance of systematic biases.
In our study, we allow for spin-induced precession in both the injection and recovery waveform model, which had not been done in previous studies.
Since spin-precession and millilensing effects share qualitatively similar signal properties (such as beats)~\cite{liu2024discern}, including spin-precession allows for a more robust exploration of the systematic biases.
Our results complement the work of~\cite{wright2024effect}, as we establish that a false positive test of GR is typically not possible due to the significant loss of SNR incurred from neglecting millilensing effects.

The remainder of this paper is organized as follows. Section~\ref{sec:waveform models} reviews the ppE and millilensing models, and discusses systematic biases and their quantification with an illustrative example.
Section~\ref{sec:results - systematic bias} presents the results of the PE study, with the injection-recovery set-up described in Sec~\ref{subsec:injections setup}, main results presented in Sec.~\ref{subsec:results}, bias characterization discussed in Sec.~\ref{subsec:quantifying_biases}, and an executive summary of key findings reviewed in Sec.~\ref{subsec:executive_summary}. 
We present analytical estimates of the systematic errors in Sec.~\ref{sec:toy_model}.
In the final Sec.~\ref{sec:conclusion}, we conclude and summarize our main findings, and we discuss possible future research.
Henceforth, we use the following conventions: bold-faced letters (typically when referring to parameters) will stand for vectors, Einstein summation convention for repeated indices on vector/matrix components (unless stated otherwise), and geometric units, in which $G=c=1$.

\section{\label{sec:waveform models}Review of waveform models and systematic bias}

In this section, we review the millilensed and ppE models, following mostly~\cite{Yunes_2009, Liu2023Exploring}. 
We also discuss the key concepts related to systematic biases arising from inaccurate models, providing an illustrative example to demonstrate how to quantify these biases in the recovered parameters.

\subsection{Waveform models}

Gravitational lensing of GWs can be modeled using various mass density profiles that correspond to the mass distribution of astrophysical objects acting as lenses. 
The simplest models include a point-mass lens (PML) and a singular isothermal sphere (SIS), parameterized by two parameters ($y, M_{Lz}$) which correspond to the (dimensionless) relative alignment of the lens and the GW source, and the redshifted lens mass. 
Both PML and SIS models assume spherically-symmetric distributions of the lens mass, and hence, may not fully capture the complexities of real astrophysical objects.
To address these complexities, in this study, we utilize the phenomenological millilensing framework of~\cite{Liu2023Exploring}. 
This framework facilitates generic testing of arbitrary lens models within the geometric optics approximation, focusing on GW observables rather than specific properties of idealized lens models.
Importantly, the phenomenological framework retains the capability to map back to simpler models, including the PML model and the SIS model, commonly used in literature.

Gravitational millilensing results in multiple signals each of which is amplitude-magnified, time-delayed, and phase-shifted compared to the original signal. 
These effects can be mathematically expressed as follows:
\begin{align}
    \begin{aligned}[c]
    \tilde{h}_L(\boldsymbol{\lambda}_U, \;\boldsymbol{\lambda}_L;f) &= {F( \boldsymbol{\lambda}_L;f)}\cdot \Tilde{h}_U(\boldsymbol{\lambda}_U;f) \\
    &= \sum_{j =1}^{K_\mathrm{max}}
    \frac{d_L}{d^\mathrm{eff}_j} \exp \left[2 \pi i f t_{j}-i \pi n_{j}\right] \tilde{h}_{U}(\boldsymbol{\lambda}_U;f). 
    \end{aligned}
    \label{eqn:lensed_waveform_short}
\end{align}
Here, $\Tilde{h}_U(\boldsymbol{\lambda}_U;f)$ is the unlensed gravitational waveform with source parameters $\BF{\lambda}_U$, and $F(\boldsymbol{\lambda}_L;f)$ is the lensing amplification function that depends on the lensing parameters $\BF{\lambda}_L = \{d^\mathrm{eff}_j, t_j, n_j, K_{\rm{max}}\}$, corresponding to the effective luminosity distance, time delay, and phase shift of the  $j^\textrm{th}$ signal, relative to the earliest-arriving signal, as well as the total number of component millilensed signals. 
The effective luminosity distance is given by $d^\textrm{eff}_j = d_L/\sqrt{\mu_{\textrm{rel},j}}$ where $d_L$ is the luminosity distance to the source, and $\mu_{\textrm{rel},j}$ is the relative magnification of the $j^\textrm{th}$ signal compared to an unlensed GW.
The resultant waveform model is the sum of these individual lensed signals and depends on the set of parameters $\BF{\lambda}_U \cup \BF{\lambda}_L$.

We can map the millilensing parameters $\BF{\lambda}_L = \{d^\mathrm{eff}_j, t_j, n_j, K\}$ to model-specific parameters. 
For a point-mass lens, which results in $K_{\rm{max}}=2$ signals, we can convert the corresponding GW observables (i.e.~the effective luminosity distance $d^\textrm{eff}_2$, and the time delay $t_2$) to the source position parameter $y$, the luminosity distance to the source $d_L$ and the redshifted lens mass $M_{Lz}$ following~\cite{takahashi2003wave} via
\begin{subequations}
\begin{align}
    t_2 &=4 M_{L z}\left( \frac{1}{2}y \sqrt{y^2+4}+\ln \frac{\sqrt{y^2+4}+y} {\sqrt{y^2+4}-y}\right), 
    \label{eq:time_delay_PML} \\
    d^\textrm{eff}_2 & = d_L \sqrt{\left|\frac{y\sqrt{y^2+4}+y^2+2}{y\sqrt{y^2+4}-y^2-2}\right|},
    \label{eq:eff_lum_dist_PML}
\end{align}    
\end{subequations}
where $M_{L z}$ is redshifted lens mass\footnote{The redshifted lens mass $M_{L z}=M_L (1+z_L)$, with $M_L$ being the (intrinsic) source-frame lens mass and $z_L$ the lens redshift.} and $d_L$ is the luminosity distance of the GW source.
For GW lensing within the geometric optics regime considered here, where wave effects can be neglected, lensed signals generically have a fixed relative phase shift.
In the case of two lensed signals, their relative phase shift is $\Delta\Phi=\pi/2$ which corresponds to a relative Morse index of $\Delta n = n_2 - n_1= 0.5$.
As we describe in Sec.~\ref{subsec:injections setup}, throughout this study, we will use the PML model as a reference model, based on which we will select three lensing configurations and convert the PML parameters (lens mass and position) to the GW observables (effective luminosity distances, arrival times, and Morse phases). 

We use the ppE framework~\cite{Yunes_2009} to describe deviations from GR to the inspiral portion of the signal. This model is described by 
\begin{equation}
    \tilde{h}_{\rm ppE}(\BF{\lambda}_{\rm GR},\beta;f) = \tilde{h}_\textrm{GR}(\BF{\lambda}_{\rm GR};f)e^{i\beta u^b}, \label{eqn:ppE_waveform}
\end{equation}
where $\tilde{h}_\textrm{GR}(\BF{\lambda}_{\rm GR};f)$ is a frequency-domain GR model with GR parameters $\BF{\lambda}_{\rm GR}$, $\beta$ is a ppE parameter that controls the magnitude of the GR deviation, and $b$ is the ppE index that controls the PN order of the ppE deviation.
Therefore, the ppE waveform model depends on the set of parameters $\BF{\lambda}_{\rm GR}\cup \beta$ for a fixed value of $b$.
In Eq.~\eqref{eqn:ppE_waveform}, we have used $u=(\pi\mathcal{M}f)^{1/3}$, which is the ``chirp orbital velocity'' with chirp mass $\mathcal{M}=\eta^{3/5} m$, total mass $m = m_1 + m_2$, and symmetric mass ratio $\eta = m_1 m_2/m^2$ for a binary with component masses $m_1$ and $m_2$.
Note that $u$ is related to the orbital velocity $v=(\pi m f)^{1/3}$ through a factor of $\eta^{1/5}$, i.e.~$u = \eta^{1/5} \, v$.
For a given $b$, the ppE waveform can be mapped to the waveform prediction in specific theories of gravity for the leading-order GR deformation in the PN approximation. 
By choosing $b$ (and thus fixing the PN order of the ppE deviation) and constraining $\beta$, a wide class of modified gravity theories can be weeded out~\cite{Yunes_2016}. 
While the ppE framework allows for both amplitude and phase deviations in the inspiral, intermediate, and merger-ringdown regime of the GW, we focus only on ppE phase deviations in the inspiral.

\subsection{Systematic bias \label{subsec:review-bias}}
\begin{figure*}[!ht]
    \centering
    \includegraphics[width=\linewidth]{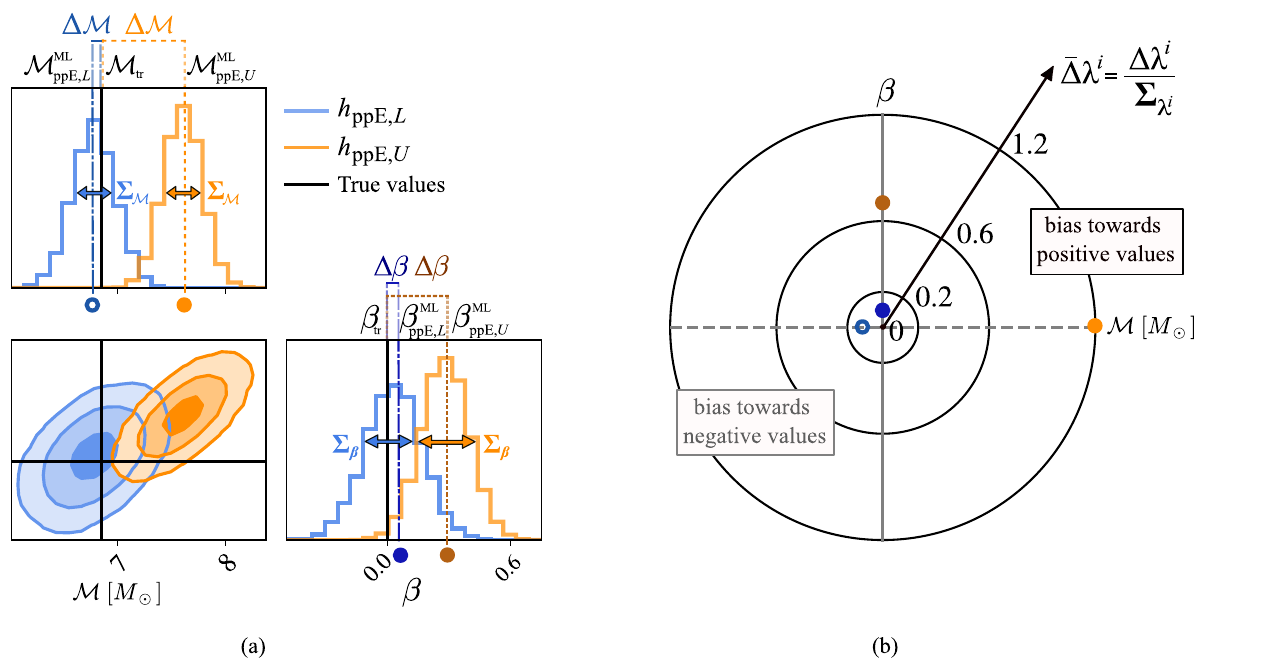}
    \caption{\justifying (a) Corner plot of parameters $\BF{\lambda}= (\mathcal{M}, \beta)$ recovered with two hypothetical models: $h_{\mathrm{ppE},L}$ and $h_{\mathrm{ppE},U}$. 
    The systematic errors $\Delta\lambda^i$, corresponding to the difference between the maximum likelihood value $\lambda_{\rm ML}^i$ (vertical dashed/dash-dotted lines) and the true value $\lambda_{tr}^i$ (black line) are annotated at the top of the one-dimensional histograms (horizontal dashed and dash-dotted lines). 
    The statistical errors $\Sigma_{\lambda^i}$, corresponding to one standard deviation of the one-dimensional posterior distributions, are annotated with arrows.
    The posterior distributions of the $h_{\mathrm{ppE},L}$ model show no significant bias, as the systematic errors are much smaller than the statistical errors.
    In contrast, the $h_{\mathrm{ppE},U}$ model shows a significant bias away from the injected values.
    (b) Spoke-wheel plot showing the ratio of the systematic error $\Delta\lambda^i$ to the statistical error $\Sigma_{\lambda^i}$, denoted as $\Bar{\Delta}\lambda^i$ for parameter $\lambda^i$. 
    The center of the plot corresponds to no deviation, and the distance from the center represents the magnitude of the normalized systematic bias.
    Filled markers in the upper right correspond to positive deviations and empty markers in the lower left correspond to negative deviations.
    In line with panel (a), the blue markers correspond to the $h_{\mathrm{ppE},L}$ model, which shows small values of the normalized systematic error ($\Bar{\Delta}\lambda^i <0.2$), with a bias in chirp mass toward negative values.
    Orange markers represent the $h_{\mathrm{ppE},U}$ model, where the systematic bias is more pronounced, with $\Bar{\Delta}\lambda^i>1$ in the chirp mass.
    }
    \label{fig:simple_example_corner}
\end{figure*}

In this section, we review some essentials that are useful when attempting to understand systematic bias. We also establish terminology and notation (following~\cite{rohit:syspaper1}), and provide an example to illustrate systematic biases in Bayesian PE due to neglecting millilensing effects.
 
Consider the hypothesis $H_L$ that the GW data is a millilensed GR signal whose waveform is described by $h_L$ with parameters $\BF{\lambda}_U \cup \BF{\lambda}_L$ (given by~\cref{eqn:lensed_waveform_short}), where $\BF{\lambda}_U$ are the (GR) source parameters and $\BF{\lambda}_L$ are the (GR) lensing parameters.
Suppose that we do not use the lensed waveform to analyze the data, but rather, we employ an unlensed model. 
Let $H_U$ be the hypothesis that the data is described by an unlensed waveform $h_{U}$ with parameters $\BF{\lambda}_U$, and $H_{\mathrm{ppE},U}$ be the hypothesis that the data is described by an unlensed ppE waveform $h_{\mathrm{ppE},U}$ with parameters $\BF{\lambda}_U \cup \{\beta\}$.
Specifically, $h_{\mathrm{ppE},U}$ reduces to $h_U$ when the ppE parameter $\beta$ vanishes, implying that the hypotheses are nested with $H_{\mathrm{ppE},U} \supseteq H_U$.
Likewise, the millilensed waveform $h_L$ is also a special case of a ppE millilensed waveform $h_{\mathrm{ppE},L}$ when the ppE parameter vanishes.
If indeed Nature is accurately described by GR, using $h_{\mathrm{ppE},L}$ to perform parameter estimation should result in the ppE parameter $\beta$ peaking at the GR value, which is zero.
In other words, \emph{injecting} a signal using $h_{\mathrm{ppE},L}$ with a vanishing ppE parameter would perfectly match the data (neglecting noise-related effects).
However, when using $h_{\mathrm{ppE},U}$, as the lensing effects are neglected in the recovery model, the estimated waveform parameters should in general be biased, affecting the resulting ppE tests of GR.

To assess systematic biases between waveform models, it is instructive to perform injection-recovery analyses as it allows one to study how biases change with the injected parameters.
Going forward, and throughout this work, we assume that the injected waveform has no specific noise realization, so as to isolate the systematic error due to waveform mismodeling from that induced by the noise realization.
We do include an averaged noise power spectral density $S_n$ to represent a Gaussian noise process, which is dependent on the detector.
We now consider an injected GR millilensed GW signal using $h_{\mathrm{ppE},L}$ with a vanishing ppE parameter ($\beta=0$), which is the same as a millilensed GW signal in GR $h_{L}$.
We perform two sets of parameter recoveries, one with $h_{\mathrm{ppE},U}$ and another with $h_U$.
For simplicity, we denote the waveform parameters that are common to the injection and recovery waveform models by $\BF{\lambda}$, which is an abstract parameter vector.
For instance, when recovering with the ppE waveform, we have $\BF{\lambda} = \BF{\lambda}_U\cup \{ \beta\}$ as the common parameters with the injected $h_{\mathrm{ppE},L}$ waveform. 
Likewise, when recovering the GR waveform we instead have  $\BF{\lambda} = \BF{\lambda}_U$ as the common parameters with the injected $h_{\mathrm{ppE},L}$ waveform.
Since the injected millilensed signal assumes GR, we know that the injected parameters are $\boldsymbol{\lambda}_{U,\rm tr} \cup \boldsymbol{\lambda}_{L,\rm tr}  \cup \{ \beta_{\rm tr}=0 \}$.
When recovering with the unlensed ppE waveform, the common injected parameters are $\boldsymbol{\lambda}_{\rm tr} = \boldsymbol{\lambda}_{U,\rm tr}  \cup \{ \beta_{\rm tr} \}$, and when recovering with unlensed GR waveforms the common injected parameters are $\boldsymbol{\lambda}_{\rm tr} = \boldsymbol{\lambda}_{U,\rm tr}$.

To quantify systematic errors, we adopt the \emph{likelihood-based systematic error} measure in waveform parameters, defined as $\Delta \BF{\lambda} = \bfml{\lambda} - \bftr{\lambda}$. 
In this measure, $\bfml{\lambda}$ are the waveform parameters of the recovery model corresponding to the maximum-likelihood point (see~\cite{Flanagan:1997kp,Cutler_2007,rohit:syspaper1} for a geometric interpretation of systematic error).
We assess the accuracy of the model by comparing the systematic error $\Delta \lambda^i$ to the statistical error $\Sigma_{\lambda^i}$, defined as 90\% confidence interval of the marginalized posterior distribution, in measuring each parameter $\lambda^i$.
The recovery model is \emph{accurate} when $\Delta \lambda^i < \Sigma_{\lambda^i}$, which is satisfied when the recovery model is nearly identical to the injected model or when the SNR is too small and there is a large statistical error in measuring the parameters.
Meanwhile when $\Delta \lambda^i > \Sigma_{\lambda^i}$, the recovery model is inaccurate, which happens when the recovery model is very distinct from the injected model or when the SNR is very large.

For pedagogical reasons and to enable an easier understanding of our conclusions, let us now introduce two visual representations of Bayesian inference results. The first representation is the usual ``corner plot'', where one shows the one- and two-dimensional marginalized posterior distribution of a set of parameters (given a model, priors on the parameters, and a signal) after the likelihood has been sufficiently well-explored through Markovian methods.  The left panel of~\cref{fig:simple_example_corner} is a simple cartoon to exemplify this type of representation for the chirp mass $\cal{M}$ and ppE parameter $\beta$ space. This figure was created by sampling 2 two-dimensional Gaussian functions, centered at some chosen values of $({\cal{M}}_{\rm ML},\beta_{\rm ML})$ and with some chosen $1\sigma$ widths $\Sigma_{\cal{M}}$ and $\Sigma_{\beta}$; it is meant to represent the results of Bayesian parameter estimation for 2 models ($h_{\mathrm{ppE},L}$ in blue and $h_{\mathrm{ppE},U}$ in orange) given some data consistent with one of them ($h_{\mathrm{ppE},L}$), and flat priors on ${\cal{M}}$ and $\beta$. Observe that the value of the ``true'' parameters (solid black lines, sometimes also known as the ``signal'' parameters or the ``injected'' parameters) $({\cal{M}}_{\rm tr},\beta_{\rm tr})$  are consistent with the posteriors obtained with the $h_{\mathrm{ppE},L}$ model, but not with those obtained with the $h_{\mathrm{ppE},U}$ model. The distances between the maximum likelihood values (which are the same as the maximum posterior values for flat priors) of the chirp mass and the ppE parameter (i.e.~$({\cal{M}}_{\rm ML},\beta_{\rm ML})$) and the injected parameters are the systematic errors $\Delta {\cal{M}} = {\cal{M}}_{\rm ML} - {\cal{M}}_{\rm tr}$ and $\Delta \beta = \beta_{\rm ML} - \beta_{\rm tr}$. Meanwhile, the statistical errors are just given by the widths of the marginalized (one-dimensional) posteriors, which were chosen in this cartoon to be $\Sigma_{\cal{M}}$ and $\Sigma_{\beta}$.

When one is interested in studying the relation between the systematic and the statistical error that results from a Bayesian parameter estimation study, a ``spokes-wheel'' plot becomes useful. This plot is a polar representation (see~\cite{rohit:syspaper1}) of the ratio of the systematic to the statistical error $\Bar{\Delta} \lambda^i\equiv \Delta \lambda^i/\Sigma_{\lambda^i}$ for every parameter in the model, a cartoon of which is shown in the right panel of Fig.~\ref{fig:simple_example_corner}. The concentric circles represent iso-ratios of the systematic to the statistical error, while the radial distance from the center represents the magnitude of the ratio\footnote{The angle of the spokes-wheel is not physically meaningful, but rather chosen to separate the errors in the different model parameters.}. Figure~\ref{fig:simple_example_corner} shows that the systematic errors induced by the $h_{\mathrm{ppE},U}$ model are much larger than those introduced by the $h_{\mathrm{ppE},L}$ model\footnote{When the injection is identical to the model, then the bias should in principle be identically zero. However, finite sampling errors and specific noise realizations may introduce small shifts in the peaks of posteriors, as we attempted to represent in the cartoon of Fig.~\ref{fig:simple_example_corner}.}, and that the systematics in the chirp mass are larger than those in the ppE parameter. 
In what will follow in this paper, we will employ these two types of figures to visually represent the real Bayesian parameter estimation results that we obtained. 

How does one quantify and characterize the systematic bias in the recovery of the ppE deviation?
We now review the systematic bias characterization outlined in~\cite{rohit:syspaper1} that helps in assessing whether one has detected a false positive deviation from GR.
Following~\cite{rohit:syspaper1}, when the systematic error in the recovery of the ppE deviation $|\Delta \beta|$ is smaller than its statistical error $\Sigma_{\beta}$, we are in the case of a ``Definite Inference of No GR Deviation''.
Meanwhile, when $|\Delta \beta| > \Sigma_{\beta}$, there is a ``Potential Inference of a False GR Deviation''.

To assess further the case of a Potential Inference of a False GR Deviation, we use both an SNR residual test and a model selection test.
The SNR residual test is controlled by the fitting factor of the ppE model $\mathrm{FF_{ppE}}$, which is simply the match between the (recovered) maximum likelihood ppE waveform and the signal.
The fitting factor essentially represents the fraction of the true SNR recovered by the ppE model.
The model selection test is controlled by the Bayes factor favoring the ppE model over the GR model, denoted by $\mathrm{BF^{ppE}_{GR}}$.
The Bayes factor is the ratio of the evidences of the ppE model and the unlensed GR model, quantifying how much the data prefers the ppE model over the unlensed GR model. 
Based on thresholds on the fitting factor of the ppE model $\mathrm{FF}_{\rm ppE}$, and the Bayes factor between the ppE and GR models $\mathrm{BF}_{\rm GR}^{\rm ppE}$, there are four possible sub-regimes to characterize a Possible Inference of a False GR Deviation.

When the $\mathrm{FF}_{\rm ppE}$ exceeds its corresponding threshold, there is negligible loss of SNR when using the ppE model.
Similarly, when $\mathrm{BF}_{\rm GR}^{\rm ppE}$ exceeds its corresponding threshold, there is a strong preference for the ppE model over the GR model.
Following~\cite{rohit:syspaper1}, we pick the fitting factor threshold as $\mathrm{FF}^{\rm dist}_{\rm ppE} = 1-N_{\rm ppE}/(2\rho^2)$, where $N_{\rm ppE}$ is the number of parameters in the ppE model and $\rho$ is the SNR of the injection.
For the model selection threshold, we pick $\mathrm{BF}_{\rm GR}^{\rm ppE} = 10$, based on the Jeffreys' scale~\cite{jeffreys1961theory}.

Based on the above thresholds, and following~\cite{rohit:syspaper1}, the four sub-regimes of a Possible Inference of a GR Deviation are the following:
\begin{enumerate}[label=\textnormal{(\Roman*)}]
    \item ``Strong Inference of No GR Deviation'':
    $|\Delta \beta| > \Sigma_{\beta}$, $\mathrm{BF}_{\rm GR}^{\rm ppE} < 10$, and 
    $\mathrm{FF}_{\rm{ppE}}<\mathrm{FF}_{\rm{ppE}}^{\rm{thresh}}$.
    \item ``Weak Inference of No GR Deviation I'': 
    $|\Delta \beta| > \Sigma_{\beta}$, 
    $\mathrm{BF}_{\rm GR}^{\rm ppE} < 10$, and $\mathrm{FF}_{\rm ppE} >\mathrm{FF}_{\rm ppE}^{\rm dist}$.
    \item ``Weak Inference of No GR Deviation II'': 
    $|\Delta \beta| > \Sigma_{\beta}$, 
    $\mathrm{BF}_{\rm GR}^{\rm ppE} > 10$, and $\mathrm{FF}_{\rm ppE} <\mathrm{FF}_{\rm ppE}^{\rm dist}$.
    \item ``Incorrect Inference of GR Deviation'': $|\Delta \beta| > \Sigma_{\beta}$, 
    $\mathrm{BF}_{\rm GR}^{\rm ppE} > 10$, and $\mathrm{FF}_{\rm ppE} >\mathrm{FF}_{\rm ppE}^{\rm dist}$.
\end{enumerate}
We refer the reader to~\cite{rohit:syspaper1} for a detailed description of what these different sub-regimes mean, although the naming scheme is self-explanatory.
We use the above classification to characterize the systematic biases inferred in this work.

Computing the systematic error in practice can be difficult due to complicated waveform templates and posterior surfaces. 
However, when the signal is loud and the systematic error due to mismodeling is small, $\Delta \BF{\lambda}$ can be computed using the LSA.
In this approximation, the waveform is linearized in $\Delta \BF{\lambda}$ about the maximum likelihood point~\cite{Flanagan:1997kp,Cutler_2007}.
For each waveform parameter $\lambda^i$, the associated systematic error $\Delta \lambda^i$ can be expressed as:
\begin{align}
    \Delta \lambda^i =C^{ij}(\bftr{\lambda}) \left(\partial_{\lambda^j} h_M(\bftr{\lambda}) | d - h_M(\bftr{\lambda}) \right),
\end{align}
where $C^{ij}(\bftr{\lambda}) = \left(\Gamma^{-1}\left( \bftr{\lambda}\right)\right)^{i j}$ is the inverse of the Fisher information matrix $\Gamma_{ij} (\bftr{\lambda}) := \left(\partial_{\lambda^i} h_M(\bftr{\lambda}) |  \partial_{\lambda^j} h_M(\bftr{\lambda}) \right)$, $d$ is the signal,
$h_{M}$ is the approximate (recovery) waveform model, the inner product is defined via
\begin{align}
    (h_1|h_2) := 4 \Re \int \frac{\tilde{h}_1(f) \tilde{h}_2^*(f)}{S_n(f)} df \,,
\end{align}
all of which are evaluated at the injected waveform parameters. 
We assume that $\Delta A =0$ (where $\Delta A$ is the error in the waveform amplitude), which is an approximation that assumes the phase modulations are more dominant than the amplitude modulations in driving the systematic bias (for further discussion, see~\cref{appendix_PE}).
In Sec.~\ref{sec:results - systematic bias}, using a nested sampling analysis, we confirm that indeed amplitude modulations play a negligible role in driving biases in phase parameters.
When the phase difference between the signal and the recovery waveform model $\Delta \Psi$ is small, linearizing in $\Delta \Psi$ results in
\begin{align}
\begin{split}
    \Delta \lambda^i &=(\Gamma^{-1}(\boldsymbol{\lambda}_{\mathrm{tr}}))^{i j} 4 \int_{f_{\min }}^{f_{\max }}\Bigg[\frac{d f}{S_n(f)} \\
    & \times A^2\left(\boldsymbol{\lambda}_{\mathrm{tr}} ; f \right)\partial_{\lambda^j} \Psi_{M}\left(\boldsymbol{\lambda}_{\mathrm{tr}} ; f \right) \Delta \Psi\left(\boldsymbol{\lambda}_{\mathrm{tr}} ; f \right) \Bigg],
    \label{eq:sys_error}    
\end{split}
\end{align}
where $\Psi_{M} $ is the phase of the approximate waveform.

For loud signals, the estimation of the Bayes factor can also be simplified using the \emph{Laplace approximation}~\cite{gregory_2005,Cornish:2011ys}, where the joint posterior is approximated as a multivariate Gaussian around the maximum posterior point.
Note that when the prior is either uniform or Gaussian, the Laplace and Fisher approximations are equivalent.
In the Laplace approximation, the Bayes factor in favor of the ppE model over the GR model is given by~\cite{Vallisneri:2012qq,Moore:2021eok,rohit:syspaper1}
\begin{align}
    \log \mathrm{BF^{ppE}_{GR}} \approx (\mathrm{FF}_{\rm ppE} - \mathrm{FF}_{\rm GR}) \rho^2 + \log O^{\rm ppE}_{\rm GR}, \label{eqn:laplace}
\end{align}
where $\mathrm{FF}_{\rm ppE}$ and $\mathrm{FF}_{\rm GR}$ are the fitting factors of the ppE and GR models respectively, while $O^{\rm ppE}_{\rm GR}$ is the Occam penalty of using the ppE model over GR,\footnote{Note that this is not the Posterior Odds (Bayes factor multiplied by the Prior Odds) sometimes with a similar notation in the literature. } and $\rho$ is the (optimal) signal SNR.
The Occam penalty $O^{\rm ppE}_{\rm GR}$ is simply the ratio of the posterior to prior volume for the additional ppE parameter.
The fitting factor can be computed in the LSA and for the ppE model (and likewise for the GR model), one obtains~\cite{rohit:syspaper1}
\begin{align}
    \mathrm{FF}_{\rm ppE} \approx \mathcal{F}(\bftr{\lambda}) + \dfrac{1}{2 \rho^2} (\Delta \lambda^i) (\Delta \lambda^j)\Gamma_{ij}(\bftr{\lambda}), \label{eqn:fitting_factor}
\end{align}
where $\mathcal{F}(\bftr{\lambda})$ is the match between the ppE model and the (millilensed) signal at the true parameters, and the second term is the correction due to the systematic error in using the approximate waveform. 
The match for an approximate waveform $h_M$ with the signal, evaluated at the true parameters $\bftr{\lambda}$, is given by
\begin{align}
    \mathcal{F}(\bftr{\lambda}) \approx \max_{\phi_c, t_c} \dfrac{(d | h_M(\bftr{\lambda}))}{\rho^2}, \label{eqn:match}
\end{align}
where the numerator is maximized over the extrinsic time and phase shift parameters of the model.
In both~\cref{eqn:fitting_factor,eqn:match}, we have made a simplification that $\rho^2 \approx (h_M(\bftr{\lambda}) | h_M(\bftr{\lambda}))$, which is valid when amplitude differences between the signal and the approximate waveform $h_M(\bftr{\lambda})$ are negligible. 
Under the LSA, it can be shown that the match (~\cref{eqn:match}) reduces to~\cite{rohit:syspaper1}
\begin{align}
    1-\mathcal{F} \approx 2\pi^2 \mathcal{N}_e^2, \label{eqn:mismatch_eff_cycles}
\end{align}
where $\mathcal{N}_e$ is the effective cycles, a noise-weighted root mean square measure of the dephasing between the signal and the approximate model $h_M$~\cite{Sampson_2014}.
The relationships between the systematic error, fitting factor, match, and effective cycles are derived in~\cite{rohit:syspaper1}.
In this paper, the approximate waveform $h_M$ is either $h_{\mathrm{ppE},U}$ or $h_{U}$.
We apply~\cref{eq:sys_error,eqn:laplace,eqn:fitting_factor,eqn:mismatch_eff_cycles} in Sec.~\ref{sec:toy_model} for a toy model to compute systematic errors when using an approximate waveform and characterize the statistical significance of the systematic biases.


\section{\label{sec:results - systematic bias} Systematic bias in parameterized tests of GR due to millilensing}

In this section, we perform an injection--recovery campaign in Bayesian parameter estimation and model selection to analyze the systematic biases due to neglecting millilensing effects. 
In Sec.~\ref{subsec:injections setup}, we first describe the millilensing and source parameter configurations used for the injections.
We then describe the different recoveries we perform using a nested sampling analysis and describe the priors and sampler settings used.
In Sec.~\ref{subsec:results}, we present the results for the systematic biases, such as the dependence of the bias on the ppE index, the lensing, and source configurations. 
In Sec.~\ref{subsec:quantifying_biases}, we characterize the systematic biases based on their statistical significance and discuss their interpretation.
We finally provide an executive summary of our main findings in Sec.~\ref{subsec:executive_summary}. 

\subsection{\label{subsec:injections setup}Injection--recovery set up}

\begin{figure*}[!ht]
    \centering
    \includegraphics[width=\textwidth]{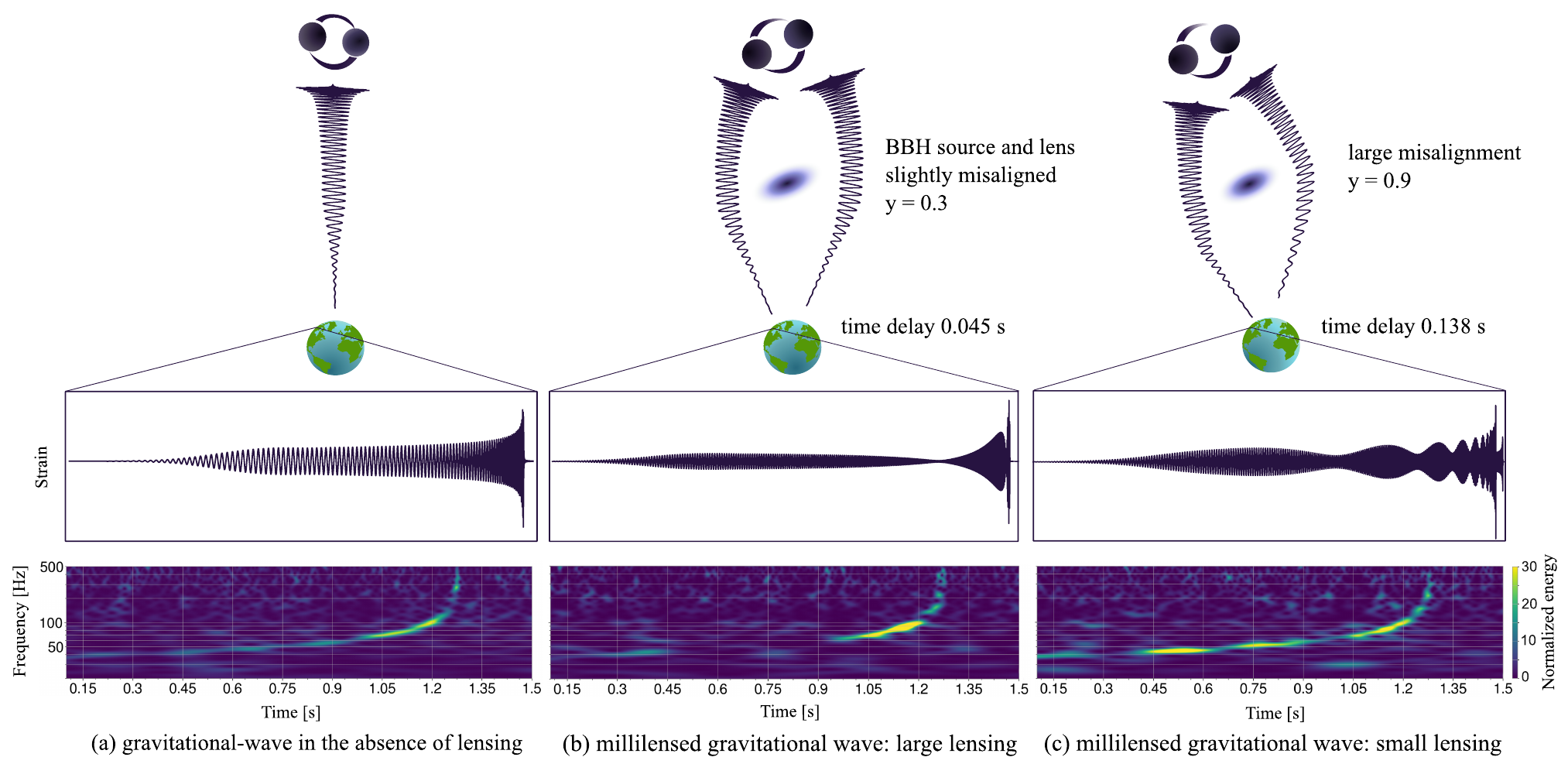}
    \caption{\justifying 
    Illustrative representation of the system setup and configurations considered: (a) GW signal in the absence of lensing, (b) millilensed GW in the large lensing scenario, with small misalignment between the BBH and the gravitational lens (small impact parameter $y=0.3$), (c) millilensed GW in the large lensing scenario, where the BBH source and the lens are largely misaligned (large impact parameter $y=0.9$).
    The middle horizontal panel displays the time-domain GW strain, while the frequency evolution over time is shown in the spectrograms in the bottom panel.
    Notable amplitude modulations induced by millilensing are observed in the waveform, appearing as \textit{beats} in the time-domain strain and as variations in energy, represented by the color intensity, in the spectrograms.
    }
    \label{fig:waveforms}
\end{figure*}

We perform an injection study of millilensed GW signals with $K_{\rm max}=2$ components, as illustrated in Fig.~\ref{fig:waveforms}.
We quantify the lensing strength by varying the relative source position parameter $y \in \{0.3, 0.6, 0.9\}$ within a point-mass lens model\footnote{
While the point-mass model is idealized and relies on unphysical assumptions, it serves as a useful reference for developing a more general phenomenological description.}.
We convert the source position parameter $y$ to the corresponding GW observables (i.e.~the effective luminosity distance $d^\textrm{eff}_2$ and the time delay $t_2$) 
of the injected millilensed signals using~\cref{eq:time_delay_PML,eq:eff_lum_dist_PML}. 
The phase shift between 2 component lensed signals in the PML model is $\Delta n=n_2-n_1 = 0.5$.
Additionally, we vary the BBH component masses $(m_1, m_2)\in \{(12, 8), (24, 16)\} M_\odot$, as well as the SNR $\in\{10, 30, 60\}$ of the injections to examine how the bias behaves when varying the properties of the detected GW signals.
We list the injection parameters used for the SNR 30 signals in~\cref{tab:injected_values}.
For signals with SNR=10 and SNR=60, we rescale the luminosity distance and the effective luminosity distance to obtain the desired SNR, while keeping all other source parameters as listed in~\cref{tab:injected_values}.
The signals are injected into ground-based detectors (LIGO Hanford, LIGO Livingston~\cite{aasi2015advanced} and Virgo~\cite{acernese2014advanced}) at the O4 sensitivity~\cite{PSD_DCC} with zero noise realization.
We use the phenomenological waveform approximant \textsc{IMRPhenomPv2}~\cite{Schmidt_2012, Hannam_2014, Khan_2019} with an injected effective spin of $\chi_\text{eff}=0.1$.
\begin{table}
\begin{ruledtabular}
    \begin{tabular}{lll}
    Symbol & Parameter & Value \\ 
    \colrule
    $(m_1, m_2)$ & BBH component masses  &  $\{(12, 8), (24, 16)\}\,M_\odot$ \\
    $d_L$ & luminosity distance  &  $\{\{774, 699, 641\},$ 
    \\
    && $\;\{1334, 1204, 1135\}\}$ Mpc \\ 
    $(a_1, a_2)$ & dimensionless spins & (0.320, 0.326) \\
    $(\theta_1, \theta_2)$ & tilt angles & (0.357, 0.403) rad \\
    $\mathrm{RA}$ & right ascension & 6.088 rad \\
    $\delta$ & declination & 0.92 rad \\
    $\psi$ & polarization & 3.16 rad\\
    $\theta_\mathrm{JN}$ & inclination & 0.519 rad \\
    $\theta_\mathrm{JL}$ & cone of precession & 3.201 rad \\
    $\phi_{12}$ & azimuthal angle & 3.574 rad \\
    $\phi$ & phase & 3.134 rad \\
    $d^\mathrm{eff}_2$ & effective luminosity  & $\{\{3127, 1262, 1534\}$, \\
    & distance & $\;\{1799, 2175, 2715\}\}$ Mpc\\
    $t_2$ & time delay &  \{0.045, 0.091, 0.138\} s\\
    $n_1$ & Morse phase & 0 \\
    $n_2$ & Morse phase & 0.5\\
    $K_{\rm{max}}$ & no. of component signals  & 2 \\
    $y$ & relative source position & $\{0.3, 0.6, 0.9\}$\\
    $M_{Lz}$ & redshifted lens mass & 3792 $M_\odot$ \\
\end{tabular}
\caption{\justifying Injected parameters of the simulated GW signals with SNR of 30. 
    The millilensing parameters were obtained from simulating three lensing setups as point-mass lenses with \textsc{lenstronomy} package \citep{birrer2018lenstronomy, Birrer2021} and then converted into GW observables $d^\mathrm{eff}_2, t_2, n_1, n_2, K_{\rm{max}}$.
    The luminosity distance and effective luminosity distance were rescaled for each lensing case to achieve the corresponding lensing magnification and an SNR of 30.}
    \label{tab:injected_values}
\end{ruledtabular}
\end{table}

We recover the signal parameters with two nested waveform models: (i) a GR waveform model without lensing effects, and (ii) a waveform model with inspiral ppE corrections added to the waveform model from (i). 
We assume \textsc{IMRPhenomPv2} as the base waveform model in both cases. 
With these two models, for a given injection, we first perform Bayesian PE, and then model selection to assess how much the non-lensed ppE waveform model is preferred over the non-lensed GR waveform model.
In the recovery of millilensed GW signals using ppE templates, we perform the recovery with one PN order correction active at a time, aiming to constrain $\beta_{\rm nPN}$.  
Specifically, we focus on the -1PN, 1PN, and 2PN orders, as these correspond to the most widely studied modified gravity theories~\cite{Yunes_2016}.

We adopt uniform priors for $\beta_{\rm nPN}$ ensuring that the contributions from terms involving $\beta_{\rm nPN}$ at a given PN order remain small relative to the corresponding GR term, which results in~\citep{rohit:syspaper1}
\begin{align}
\vert \beta_{(5+b)\rm PN} \vert < \dfrac{3}{128} (\pi \mathcal{M} f)^{-(5+b)/3}. \label{eqn:ppE_prior}
\end{align}
For the $\{-1, 1, 2\}$ PN orders,~\cref{eqn:ppE_prior} results in the prior bounds $\{\pm3.14\times 10^{-4}, \pm1.41\times 10^{-1}, \pm8.43\times 10^{-1}\}$ for $\{\beta_{-1\rm PN},\beta_{1\rm PN},\beta_{2\rm PN}\}$ respectively (given the prior range on $\mathcal{M}$ and the frequency window used in our analysis).
In Table~\ref{tab:signal_parameters}, we list the priors used for all the parameters of the recovery model.
The luminosity distance prior was set to uniform for all injections, with a lower bound of 100 Mpc and the upper bound varying between 2000 Mpc and 4000 Mpc, depending on the injection value (see~\cref{tab:injected_values}).

\begin{table}
\begin{ruledtabular}
\resizebox{0.97\linewidth}{!}{%
\begin{tabular}{lccl}
Symbol & Parameter & Prior & Range \\
\colrule
$\mathcal{M}$ [$M_\odot$] & chirp mass & uniform & (5, 20), (5, 40) \\
$q$ & mass ratio &  uniform & (0.125, 1) \\
$a_1, a_2$ & dimensionless spins & uniform & (0, 0.99) \\
$\theta_1, \theta_2$ [rad] & tilt angles & sine & $(0, \pi)$\\
$d_L$ [Mpc] & luminosity distance & uniform & varied (see text)\\
$\theta_{JN}$ [rad]& inclination angle & sine & $(0, \pi)$ \\
$\phi_{JL}$ [rad]&  cone of precession & uniform & $(0, 2\pi)$ \\
$\psi$ [rad]&  polarisation angle & uniform & $(0, \pi)$ \\
RA [rad]  & right ascension & uniform & $(0, 2\pi)$\\
DEC [rad] & declination & cosine & $(-\pi/2, \pi/2)$ \\
$\phi_c$ [rad] & phase at coalescence & uniform & $(0, 2\pi)$ \\
$t_c$ [s] & coalescence time & uniform & $(t_c^\textrm{inj}-0.2, t_c^\textrm{inj}+0.2)$ \\
$\beta_\textrm{ppE}$ & ppE phase deviation & uniform & (see text)
\end{tabular}
}
\caption{\justifying Parameters describing the parameters of the recovery models we employ for Bayesian parameter estimation, together with their corresponding priors.}
\label{tab:signal_parameters}
\end{ruledtabular}
\end{table}
Given this suite of injections and these two recovery models, we explore the likelihood (assumed to be stationary and Gaussian)
with the Bayesian inference library \textsc{bilby}~\cite{Ashton2018Bilby:Astronomy} 
through nested sampling \textsc{dynesty} sampler~\cite{skilling2006nested}, using 2000 live points.
We performed robustness checks by varying the nested sampler settings. In particular, we increased the number of live points from 2000 to 4096, changed the value of the number of autocorrelation times $\texttt{nact}\in\{5, 10, 20, 30\}$, decreased the stopping criteria \texttt{dlogz} from 0.1 to 0.01, and used various sampling seeds to ensure that our results are robust.
The tests led to consistent results, and we proceeded with our analysis with the optimal settings of 2000 live points, an \texttt{nact} of 10 and a \texttt{dlogz} of 0.1.
The sampling was performed across all source parameters, with source masses parameterized by the chirp mass $\mathcal{M}$ and the mass ratio $q=m_2/m_1$.

\subsection{Systematic Bias Results of Bayesian Analysis}
\label{subsec:results}

\begin{figure*}
    \centering
    \includegraphics[width=
    \linewidth]{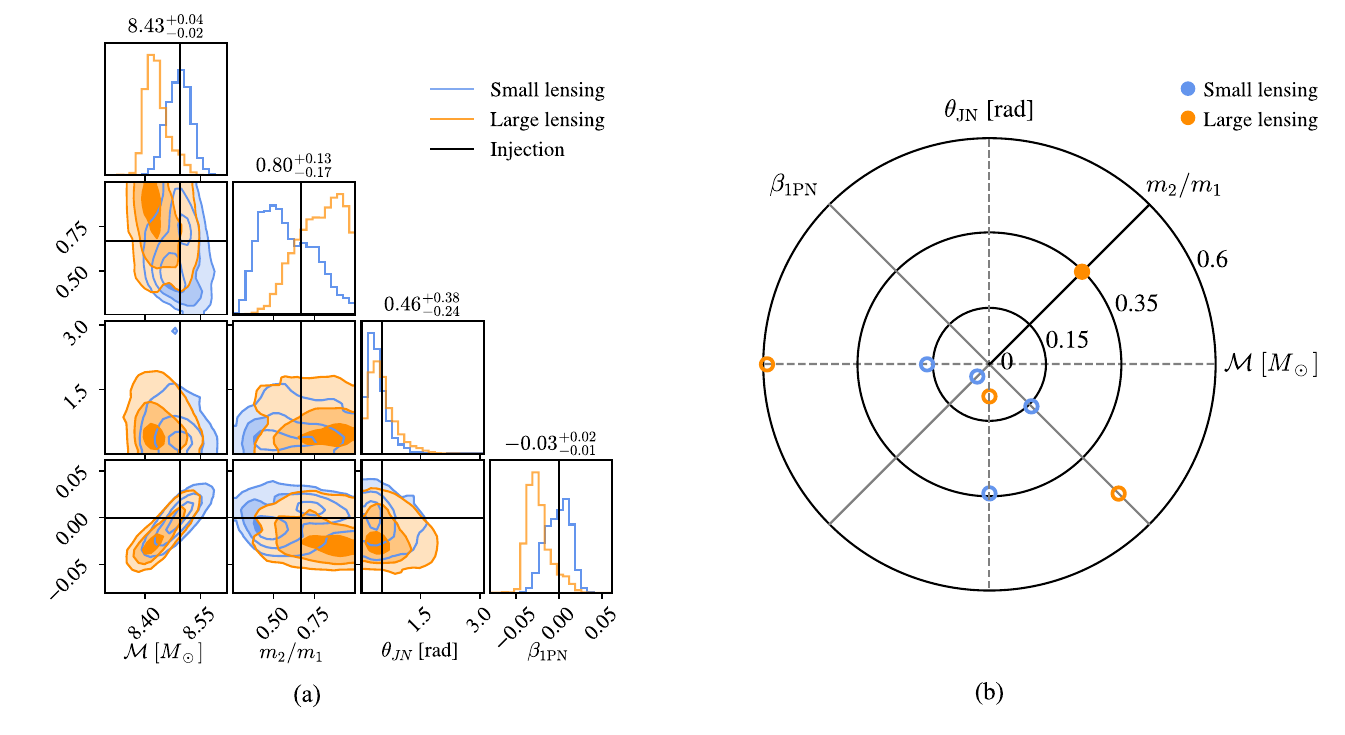}
    \caption{ \justifying Comparison of small-lensing (large impact parameter, $y=0.9$) vs large-lensing (small impact parameter, $y=0.3$) recovery of chirp mass $\mathcal{M}$, mass ratio $m_2/m_1$, inclination angle $\theta_\text{JN}$ and $\beta_\mathrm{1PN}$, for a $(12, 8)M_\odot$ system with an injected SNR of 30 and recovered with ppE model with 1PN phase correction:
    (a) corner plot comparing recoveries for the two lensing strengths, where the numerical values at the top of each column correspond to the large-lensing results.
    (b) wheel plot showing the normalized systematic error of the parameters.
    The large-lensing scenario shows a noticeable bias, particularly in the chirp mass and $\beta_\mathrm{ppE}$ distributions.}
    \label{fig:corner4d_small_large_lens}
\end{figure*}

With the injections set up, we now present our results, which address the key research questions presented in Sec.~\ref{sec:introduction}.
First, we examine the systematic bias in the recovered parameters that results when using the ppE model.
In the large lensing case, the chirp mass $\mathcal{M}$ exhibits the most significant bias among all the intrinsic parameters (see Fig.~\ref{fig:corner4d_small_large_lens}).
Since the chirp mass is correlated with $\beta_\textrm{nPN}$, a comparable systematic error is also observed in the recovery of $\beta_\textrm{nPN}$ (see Fig.~\ref{fig:corner4d_small_large_lens}~(a)). 
In the small lensing scenario, the inclination angle $\theta_{JN}$ appears to be the most biased parameter, as shown in Fig.~\ref{fig:corner4d_small_large_lens}~(b).
For a precessing signal, the inclination exhibits bias even when the recovery model matches the injection model.
Interestingly, the bias in $\theta_{JN}$ is reduced with larger lensing, which can be attributed to the amplitude modulation controlling the biases in inclination and distance, while the phase modulation primarily influences biases in intrinsic parameters.

\begin{figure*}
    \centering
    \includegraphics[width=0.9\linewidth]{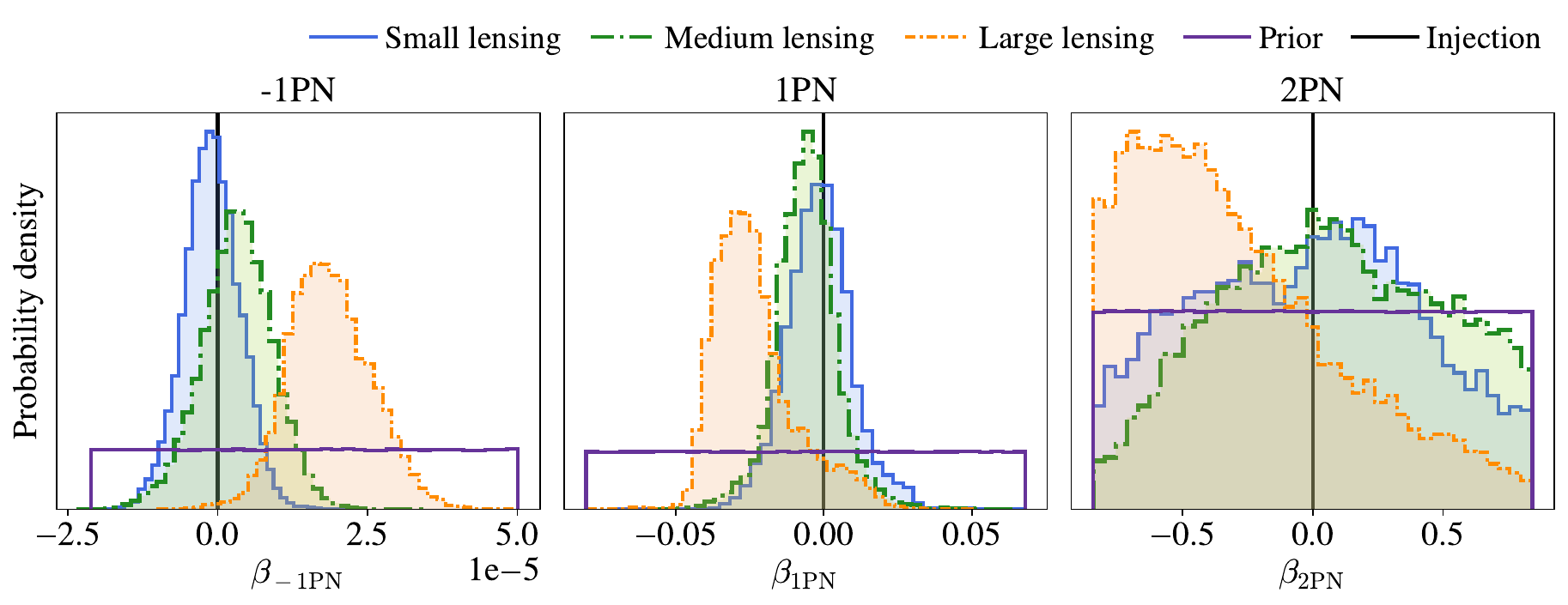}%
    \caption{ \justifying Recovery of $\beta_{\rm nPN}$ at -1PN, 1PN and 2PN with varying lensing scenarios: small lensing $y=0.9$ (blue), medium lensing $y=0.6$ (green) and large lensing $y=0.3$ (orange) for a small mass system $(12, 8)M_\odot$ and injected SNR 30. The large lensing cases show most bias towards positive values at -1PN order and towards negative values at 1PN and 2PN orders.
    }
    \label{fig:mchirp_beta_vary_lens_combined}
\end{figure*}

Second, we determine the lensing strength that can lead to significant biases i.e., when the systematic error exceeds the statistical error, and identify the PN orders at which these biases are most pronounced. 
From the recovery of $\beta_{\rm nPN}$, we find that the large lensing scenario ($y=0.3$) induces significant biases, particularly at the -1PN order (see Fig.~\ref{fig:mchirp_beta_vary_lens_combined}).
Notably, the injection values fall outside the 90\% credible interval (CI) of the recovered posterior distribution.
Furthermore, we note that the size of lensing affects the extent of systematic biases.
In the medium- and small-lensing cases, the shifts in the recovered distributions relative to the injected values are minimal, with the injection values remaining within the 90\% CI, consistent with a GR recovery.
Notably, both parameters exhibit biases towards positive values for the -1PN recovery, while showing biases towards negative values for the 1PN and 2PN orders. 
This tendency can be explained analytically, as discussed in Sec.~\ref{sec:toy_model}.

\begin{figure}[!ht]
    \centering
    \includegraphics[width=\columnwidth]{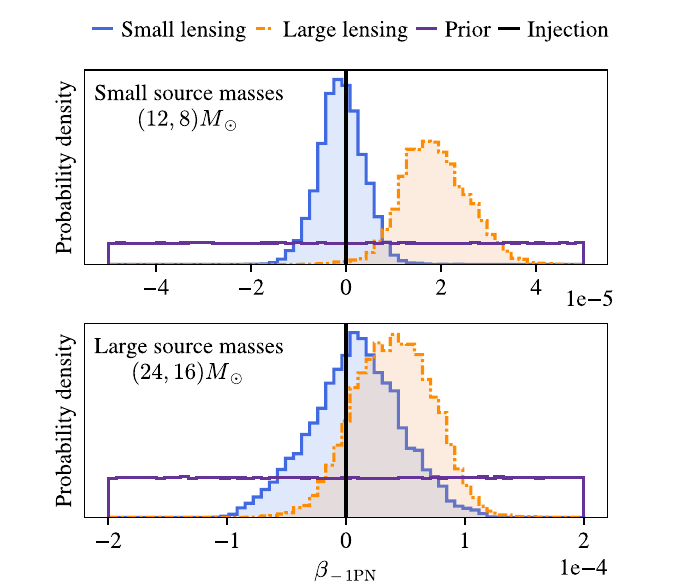}
    \caption{ \justifying Comparison of $\beta_{\rm{-1PN}}$ recovery for small source masses (upper panel) and large source masses (bottom panel), in the large-lensing (orange) and small-lensing (blue) scenarios with an injected SNR of 30.
    The injection is indicated by the black vertical line, and the prior is denoted by the purple histogram.
    Observe that the injection has very low posterior support for the small mass system, compared to the large mass system.
    Despite the absolute systematic error increasing with total mass (see also~\cref{tab:sampling_results_small_masses,tab:sampling_results_large_masses}), the ratio of systematic to statistical error in $\beta_{\rm -1PN}$ is smaller for the larger mass system, owing to the wider posteriors (a result of fewer cycles in band).
    }
    \label{fig:large_vs_small_mass_mchirp_beta}
\end{figure}

The results presented above apply to a GW source with a total mass of $m=20 M_\odot$.
This choice is motivated by the expectation that lensing effects are more pronounced in small total mass systems, potentially leading to larger biases.
To further investigate the relationship between systematic bias and the system's total mass, and hence signal duration, we compare the recovered chirp mass $\mathcal{M}$ and $\beta_\textrm{nPN}$ for two sets of GW signals: one with $m=20 M_\odot$ (long duration) and another with $m=40 M_\odot$ (short duration).
For the small lensing case with $y=0.9$, both the small and large mass systems result in insignificant systematic errors at the same SNR and same PN order, as shown by the blue histograms in~\cref{fig:large_vs_small_mass_mchirp_beta}.
When the lensing strength is increased to $y=0.3$, for the -1PN recovery, the systematic bias is more significant when $m = 20 \msun$, than when $m=40 \msun$.
We note that while the ratio $\vert \Delta\beta/\Sigma_{\beta} \vert$ is larger for the $m = 20 \msun$ case, the absolute systematic error $\vert \Delta \beta \vert$ is instead larger for the $m=40 \msun$ case.
The reason for this is due to the larger statistical error for a $m=40 \msun$ source, owing to the fewer GW cycles it spends in the detector's sensitivity band, as compared to the $m=20 \msun$ source.
Thus, our observation supports the expectation that lensing effects accumulate over the signal duration, leading to more significant systematic biases in smaller-mass signals at a fixed SNR and PN order.

\begin{figure}
    \centering
    \includegraphics[width=\linewidth]{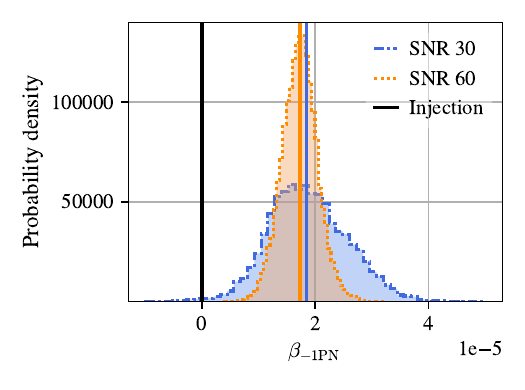}
    \caption{ \justifying 
    Recovery of $\beta_{\rm{-1PN}}$ for a small mass $(12, 8)M_\odot$ system in the large lensing $y=0.3$ case with injected SNR of 30 (blue) and 60 (orange).
    Observe that there is significant systematic error at an SNR of 30, which becomes even more pronounced at a higher SNR of 60.
    The blue and orange vertical lines represent the recovered maximum likelihood points of the two distributions, while the black vertical line is the injection.
    The systematic bias is evident at an SNR of 30, as the SNR increases, statistical errors decrease, making the systematic bias even more pronounced for the higher SNR signal.
    }
    \label{fig:beta_vs_SNR}
\end{figure}

Next, we investigate how systematic and statistical errors change with increasing SNR of the injected signal.
For the $(12,8)\msun$ system with large lensing $y=0.3$, we
perform an injection at an SNR of 60, in addition to the previously discussed SNR of 30.
As illustrated in the -1PN case in Fig.~\ref{fig:beta_vs_SNR}, the recovery of
$\beta_{-1PN}$ becomes increasingly biased with higher SNR due to the decreasing statistical error.
Notably, the injected $\beta_\text{-1PN}=0$ falls outside of the 90\% CI even at an SNR of 30, as shown in~\cref{fig:beta_vs_SNR}.
We present the high SNR recoveries of chirp mass and $\beta_{\rm{nPN}}$ at -1PN, 1PN, and 2PN orders in~\cref{fig:corner_snr30_Vs_snr60_combined}, where notably at higher PN orders, the injection value remains within 90\% CI of posterior distributions. 
The observed trend across different PN orders is related to the correlation between $\beta_{\rm nPN}$ and $\mathcal{M}$, which is strongest at -1PN and decreases with higher PN orders as illustrated in the 2D correlation plots in the left bottom panel of each corner plot in~\cref{fig:corner_snr30_Vs_snr60_combined}.

\subsection{Characterization and interpretation of the biases \label{subsec:quantifying_biases}}

We characterize the systematic biases to assess which types of signals may lead to ``Incorrect Inferences of a GR deviation'' (i.e.~false positive detections of GR deviations), based on our discussion in Sec.~\ref{subsec:review-bias}.
We compute the Bayes factor in favor of the ppE model over the GR model, denoted as $\mathrm{BF}^\text{ppE}_\text{GR}$, by taking the ratio of the evidences (obtained via nested sampling) for each model.
The error in the evidences of each model thus propagates to an error in $\mathrm{BF}^\text{ppE}_\text{GR}$, and this error is essentially controlled by what \texttt{dlogz} is used for the runs.
Given our convergence checks by varying \texttt{dlogz}, the estimates on $\mathrm{BF}^\text{ppE}_\text{GR}$ are robust and the errors from the evidence calculation do not affect the classification of the systematic biases.
For the fitting factor of the ppE model, denoted by $\mathrm{FF}_{\rm ppE}$, we compute the match between the maximum likelihood ppE model and the injected signal.
The error from estimating the maximum likelihood, which is essentially a sampling error, is what affects the accuracy of the fitting factor estimate the most.
Given the robustness of our sampling, the fitting factor estimate is also robust.

In~\cref{tab:sampling_results_small_masses,tab:sampling_results_large_masses}, we present our estimates for the systematic error $\Delta \beta$, systematic error normalized by the statistical error $(\Delta \beta/\Sigma_{\beta})$, fitting factor $\mathrm{FF}_{\rm ppE}$, and Bayes factor $\mathrm{BF}^\text{ppE}_\text{GR}$ for the injection--recoveries performed at SNR of 30.
For the $(12,8)\msun$ system (\cref{tab:sampling_results_small_masses}), at a given lensing strength, the systematic error in $\Delta \beta$ increases with increasing PN order.
Meanwhile, at a given PN order, the systematic error and preference for the ppE model increases with lensing strength.
In particular, we obtain $\mathrm{BF}^\text{ppE}_\text{GR}>1$ for all three PN orders in the large lensing scenario. 
Similarly, we observe a decreasing trend in $\mathrm{FF}_{\rm{ppE}}$ with increasing lensing strength, with significant SNR loss in the large lensing case, where $\mathrm{FF}_{\rm{ppE}}\simeq0.81$ for both 1PN and 2PN. 
For the large mass system, the trends are less pronounced than for smaller masses (see also~\cref{fig:BF}); nonetheless, the highest values of $\mathrm{BF}^\text{ppE}_\text{GR}$ are around 1.5, indicating a preference for the ppE model, although this is not significant. 
With these classifications, we can categorize the PE results, as shown in the last columns of~\cref{tab:sampling_results_small_masses,tab:sampling_results_large_masses}.
Notably, the FF is consistently below the threshold across all cases studied, excluding the incorrect inference of GR deviation.
As shown in~\cref{tab:sampling_results_small_masses}, for the large lensing case, the BF at all PN orders studied falls within the range $1<\mathrm{BF}^\text{ppE}_\text{GR}<10$, suggesting a potential deviation from GR. 
However, the FF for these results remains below the threshold, indicating that the ppE model is not a good fit to the injected data, and the observed nonzero deviation could thus arise due to mismodeling.
Thus, even for the large lensing cases, the results indicate a strong inference of no GR deviation for the signals studied.

\begin{table*}
\begin{ruledtabular}
\begin{tabular}{lrrrrrr}
PN  & lensing & $\Delta\beta$ & $\vert\Delta \beta/\Sigma_\beta\vert$ &  $\text{FF}_\text{ppE}$ & $\text{BF}^\text{ppE}_\text{GR}$ & Classification \\
\colrule
    -1 & small  & $4.033\times 10^{-6}$ & 0.529 & 0.922 & 0.006  &  Definite no GR deviation \\
   & medium & $3.845\times 10^{-6}$ & 0.392  & 0.876 & 0.014 &  Definite no GR deviation \\
   & large  & $1.495\times 10^{-5}$ & 1.343 & 0.817 & 2.098  &  Strong no GR deviation \\
\colrule 
1 & small   & $6.029 \times 10^{-4}$  & 0.033 & 0.922 & 0.087 &  Definite no GR deviation\\
  & medium  & $-3.141 \times 10^{-3}$ & 0.050 & 0.875 & 0.203 &  Definite no GR deviation \\
  & large   & $-2.273 \times 10^{-2}$ & 0.969 & 0.813 & 1.247 & Definite no GR deviation$^*$ \\
\colrule
2 & small   & $-1.542 \times 10^{-1}$ & 0.279 & 0.922 & 0.320 & Definite no GR deviation \\
  & medium  & $2.583 \times 10^{-1}$ & 0.192 & 0.875 & 1.157  & Definite no GR deviation \\
  & large   & $-5.083 \times 10^{-1}$ & 1.653 & 0.813 & 4.358 & Strong no GR deviation \\

\end{tabular}
\caption{\justifying
PE sampling results obtained for a small BBH mass system $(12, 8)\msun$ and injected SNR of 30.
For a given PN order and lensing strength, we obtain the systematic error $\Delta\beta$, the ratio of the systematic to statistical error $\vert\Delta \beta/\Sigma_\beta\vert$, the fitting factor $\text{FF}_\text{ppE}$, and the Bayes factor $\text{BF}^\text{ppE}_\text{GR}$. 
Based on these quantities, we classify the results into one of the four sub-regimes of a possible inference of GR deviation.
All results fall within either region of Definite Inference of No GR Deviation or Strong Inference of No GR Deviation, suggesting that the biases observed in PE results are due to model inaccuracies.
For the 1PN recovery in the large-lensing case, we note that the results are at the boundary between the Definite and Strong Inference of no GR Deviation, with $\vert\Delta \beta/\Sigma_\beta\vert \simeq 1$ and the Bayes factor indicating weak preference for the ppE model over the GR model.
}
\label{tab:sampling_results_small_masses}
\end{ruledtabular}
\end{table*}

\begin{table*}
\begin{ruledtabular}
\begin{tabular}{lrrrrrr}
PN  & lensing & $\Delta\beta$  & $\vert\Delta \beta/\Sigma_\beta\vert$ &  $\text{FF}_\text{ppE}$ & $\text{BF}^\text{ppE}_\text{GR}$ & Classification \\
\colrule
-1 & small  & $1.676\times 10^{-5}$ & 0.259 & 0.922 & 0.117  &  Definite no GR deviation \\
   & medium & $ 3.877\times 10^{-5}$  & 0.582 & 0.876 & 0.070 &  Definite no GR deviation \\
   & large  & $ 5.491 \times 10^{-5}$ & 0.949 & 0.816 & 0.150 &  Definite no GR deviation \\
\colrule 
1 & small  & $-5.860\times 10^{-3}$ & 0.150 & 0.922 & 0.421 & Definite no GR deviation \\
  & medium & $-5.899\times 10^{-3}$ & 0.161 & 0.875 & 1.360 & Definite no GR deviation \\
  & large  & $-7.034 \times 10^{-2}$ & 1.550 & 0.818 & 1.22 & Strong no GR deviation \\
\colrule
2 & small  & $-3.719\times 10^{-1}$ & 0.586 & 0.922 & 1.41 & Definite no GR deviation \\
  & medium & $-2.844\times 10^{-1}$ & 0.493 & 0.876 & 1.48 & Strong no GR deviation \\
  & large  & $-5.503 \times 10^{-1}$ & 1.417 & 0.815 & 0.589 & Definite no GR deviation  \\

\end{tabular}
\caption{\justifying 
PE sampling results obtained for a large BBH mass system $(24, 16)\msun$ and injected SNR of 30.
Same as in~\cref{tab:sampling_results_small_masses}, the columns correspond to the PN order, lensing strength, the systematic error $\Delta\beta$, the ratio of the systematic to statistical error $\vert\Delta \beta/\Sigma_\beta\vert$, the fitting factor $\text{FF}_\text{ppE}$, and the Bayes factor $\text{BF}^\text{ppE}_\text{GR}$.
}
\label{tab:sampling_results_large_masses}
\end{ruledtabular}
\end{table*}

\subsection{Executive summary of the results}
\label{subsec:executive_summary}

\begin{figure}
    \centering
    \includegraphics[width=\linewidth]{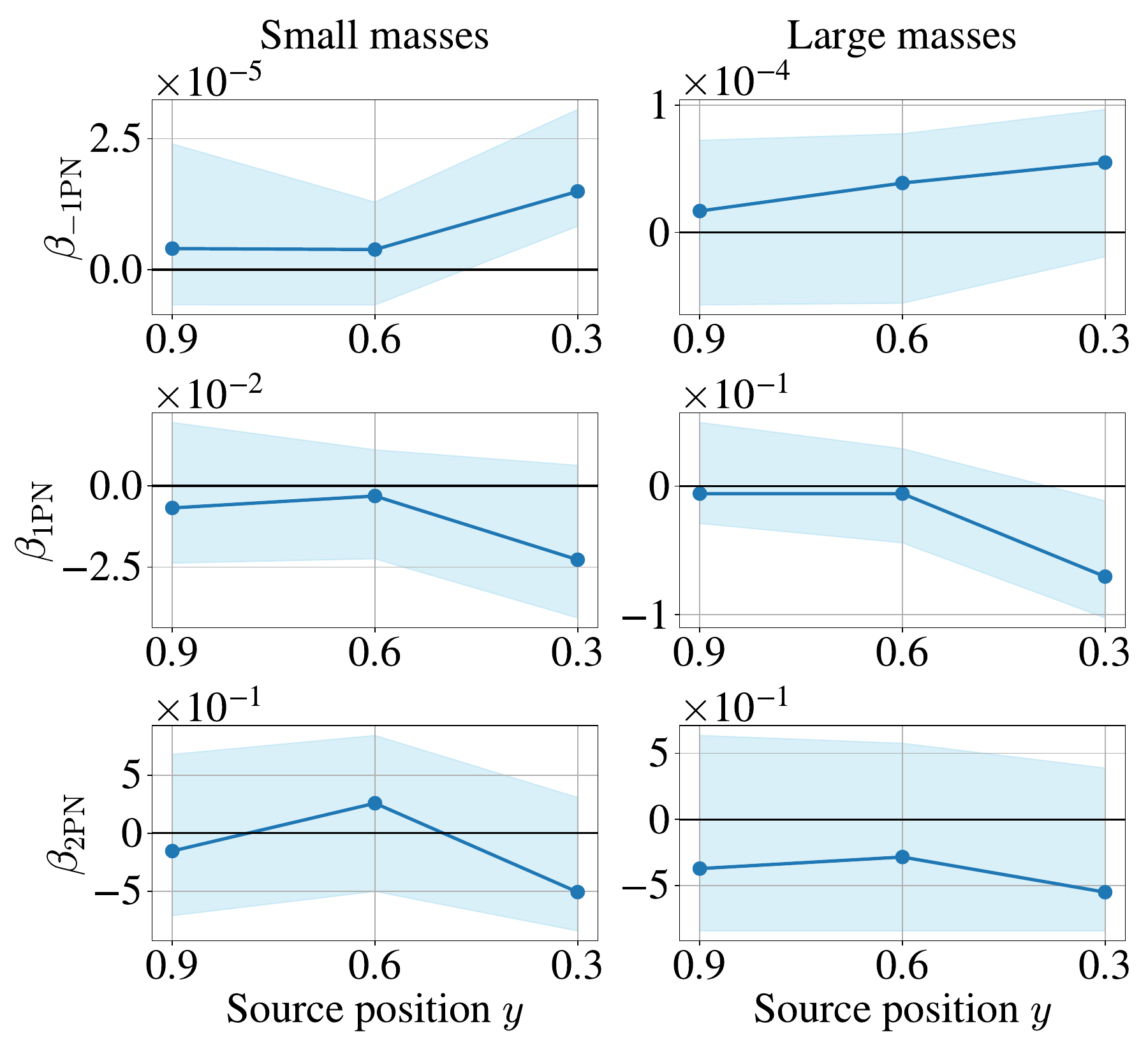}
    \caption{\justifying Systematic biases on $\{\beta_\mathrm{-1PN}, \beta_\mathrm{1PN}, \beta_\mathrm{2PN}\}$ for the small-mass $(12, 8)M_\odot$ (left column) and large-mass $(24, 16)M_\odot$ system (right column) with injected SNR 30. 
    The source position $y\in\{0.9, 0.6, 0.3\}$ corresponds to small, medium, and large lensing, respectively.
    The solid blue lines correspond to the maximum likelihood values evaluated at the three corresponding source position values, and the shaded regions represent 90\% CIs.
    As can be noted, the -1PN order shows an increasing trend with decreasing source position (increasing lensing strength), contrary to 1PN and 2PN orders which decrease with decreasing source position.
    The small- and large-mass systems differ in the order of magnitude of $\beta_\text{-1PN}$ and $\beta_\text{1PN}$.
    }
    \label{fig:beta_vs_y_large_mass}
\end{figure}

\begin{figure}
    \centering
    \includegraphics[width=\linewidth]{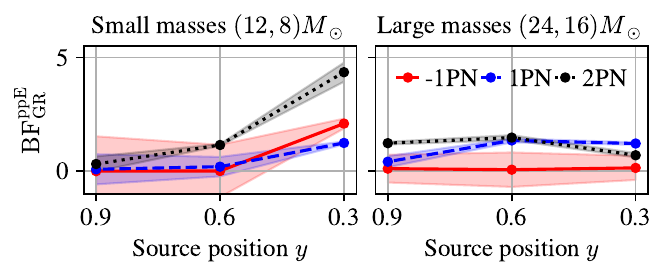}
    \caption{ \justifying
    Bayes factor comparing the ppE and GR models, obtained from the PE of the small source mass $(12, 8) M_\odot$ (left)  and large source mass $(24, 16) M_\odot$ system (right), with ppE recoveries at -1PN, 1PN and 2PN orders with shaded regions representing the uncertainties.
    The horizontal axis represents the relative source position, corresponding to small lensing ($y=0.9$), medium lensing ($y=0.6$), and large lensing ($y=0.3$).
    For the small masses (left) there is an observable trend of BF increasing with increasing lensing strength. 
    For the large masses (right) no clear trend is observable, yet all values remain within $\mathrm{BF}^\text{ppE}_\text{GR}<2$ indicating no strong preference for one model over another.
    }
    \label{fig:BF}
\end{figure}

\begin{enumerate}[leftmargin=*]
    \item \textbf{Which parameters are biased the most?} 
    In the large-lensing scenario, we find that the chirp mass $\mathcal{M}$ and $\beta_\textrm{ppE}$ exhibit the most significant biases (see ~\cref{fig:corner4d_small_large_lens}). 
    In the small-lensing case, we also observe a bias in the inclination angle $\theta_{JN}$, demonstrating the inaccuracy of the unlensed model affecting the extrinsic parameters. 
    Although these biases might not significantly impact astrophysical inference, they indicate that the model's inaccuracies are substantial enough to influence inference related to tests of GR.
    \item \textbf{What lensing size causes biases and at which PN orders?} 
    For a small mass $(12, 8) M_\odot$ system, the -1PN, 1PN, and 2PN ppE recovery models are significantly biased in the large lensing scenario when the GW source and the lens are highly misaligned ($y=0.3$) (see~\cref{tab:sampling_results_small_masses}).
    \item \textbf{How does the bias change with lensing size?} 
    The bias increases significantly with the strength of lensing. 
    For small lensing, the normalized systematic error for $\beta_\textrm{1PN}$ is approximately 0.15. 
    In contrast, for large lensing cases, this error increases to about 0.5, with very limited support for GR (see ~\cref{fig:corner4d_small_large_lens,fig:beta_vs_y_large_mass}).
    
    \item \textbf{At what SNR is the bias significant?} 
    As the SNR increases, the statistical error decreases, making the systematic biases more apparent for higher SNR signals.
    Comparing signals and recoveries at SNRs of 10, 30, and 60, we find the following:
    \begin{itemize}
        \item For the -1PN recovery model, the injected value of $\beta_{\rm nPN}$ falls outside the recovered posterior at SNR 30 (left panel in~\cref{fig:beta_vs_SNR}).
        \item For the 1PN and 2PN recovery models, the biases in $\beta_{\rm nPN}$ recovery are smaller, with the injected value positioned at the edge of the posterior even at SNR 60, as illustrated in~\cref{fig:beta_vs_SNR}.
    \end{itemize}

    \item \textbf{How do systematic biases change with source mass (signal duration)?}
    The absolute systematic error increases with total mass, fixing the lensing strength and PN order of ppE deviation.
    There is no clear trend in the significance of the systematic bias (ratio of systematic to statistical error) due to how the statistical error changes with varying total mass, lensing strength, and PN order of ppE deviation (see~\cref{tab:sampling_results_small_masses,tab:sampling_results_large_masses}).
    \item \textbf{How can we interpret and quantify the biases and when are they significant?}
    We characterize the statistical significance of the systematic biases with the Bayes factor and fitting factor.
    For millilensed signals with SNR of 30 and the strongest lensing effect ($y=0.3$), we find that the (unlensed) ppE model is weakly favored over the (unlensed) GR model (see~\cref{fig:BF}). 
    We also observe a significant loss of SNR in the ppE model recovery, which increases with the lensing strength (\cref{tab:sampling_results_small_masses,tab:sampling_results_large_masses}). This implies that a false Bayes-factor inference of a GR deviation would be avoided through an SNR loss (or FF) test.
\end{enumerate}

\section{\label{sec:toy_model}
Phenomenological explanation of systematic bias results}

\begin{figure}[h]
    \centering
    \includegraphics[width=0.5\textwidth]{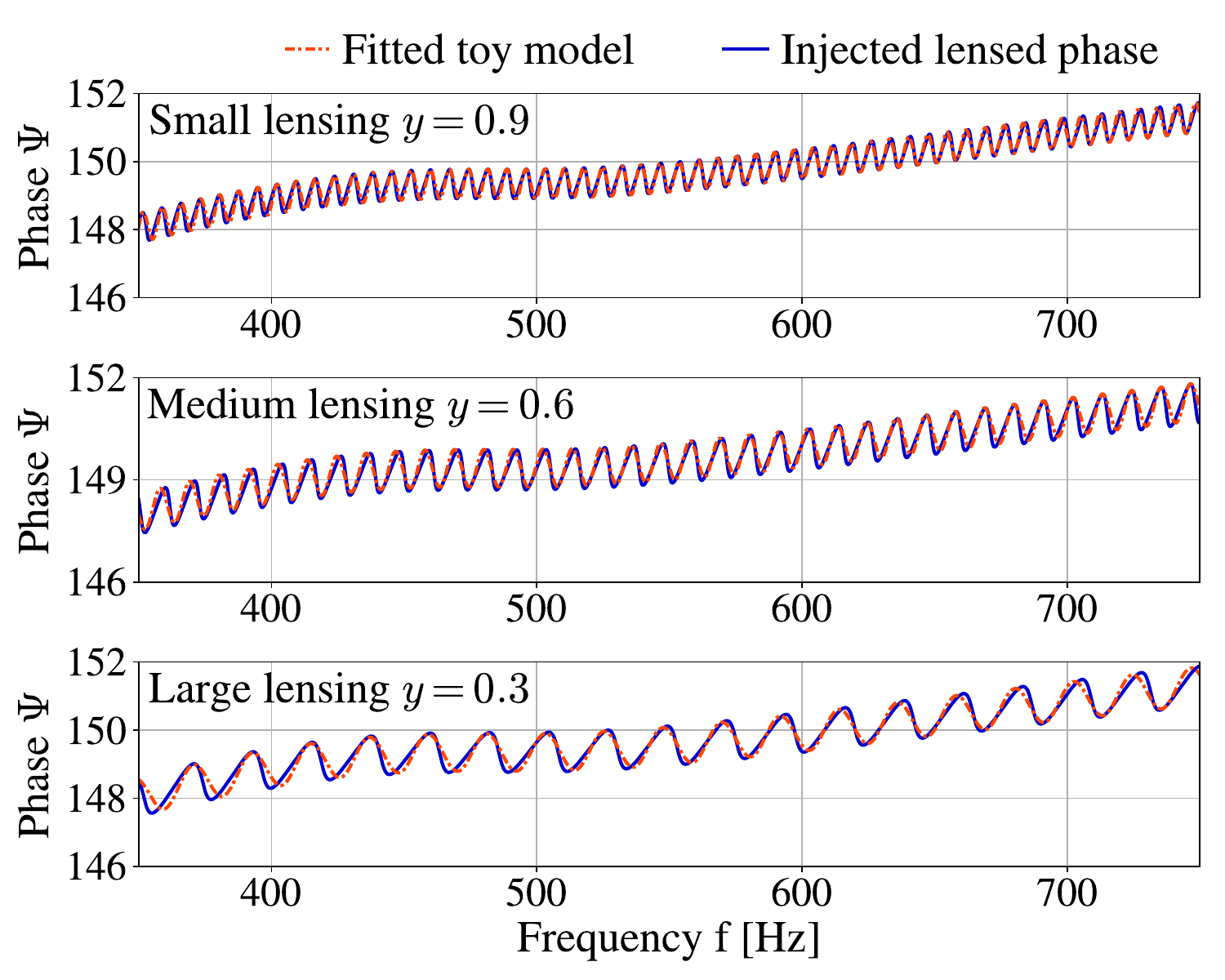}
    \caption{\justifying Comparison of the GW phase constructed with toy model fitted against actual injected GW phase for a large source mass ($M_\text{tot}=40M_\odot$), small lensing $y=0.9$ case (\textit{top}), medium lensing $y=0.6$ (\textit{middle}) and large lensing $y=0.3$ case (\textit{bottom}). The corresponding fitted parameters are listed in~\cref{tab:toy_model_fits}.}
    \label{fig:toy_model_large_masses_fit}
\end{figure}

In order to better understand the dependence of the bias on the ppE index and the lensing strength (shown in~\cref{fig:mchirp_beta_vary_lens_combined}), we develop a toy model for millilensed signals and use the LSA to estimate systematic errors semi-analytically.
Doing so allows us to obtain analytic insight into the systematic biases that result from neglecting millilensing effects.
We construct our toy model for millilensing by only modeling the phase modulations, since amplitude modulations are sub-dominant in driving biases in the phase parameters (see~\cref{fig:AM_PM}).
As shown in~\cref{fig:toy_model_large_masses_fit}, the phase modulations due to millilensing are oscillatory in frequency.
To capture this oscillatory behavior, we express the phase (evaluated at the injected parameters) of the toy model as
\begin{align}
    \Psi_\mathrm{milli}(\bftr{\lambda}; f) &= \Psi_\mathrm{GR}(\bftr{\lambda}; f) + A\sin(Bf), 
    \label{eq:toy_model_phase}
\end{align}
where $ \Psi_\mathrm{GR}(\lambda_\mathrm{tr}; f) = (3/128) u_{\rm tr}^{-5}$ with $u_{\rm tr} = (\pi \mathcal{M}_{\rm tr} f)^{1/3}$, and $\bftr{\lambda} = \{ \mathcal{M}_{\rm tr},0\}$. 
Here, $A$ and $B$ represent the amplitude and period of the lensed dephasing, respectively.
These parameters depend on the mass of the BBH system and the lensing configuration parameters, including the relative source position $y$ and the number of component lensed signals $K$.
For the three millilensing configurations (small, medium, and large) described in Sec.~\ref{sec:waveform models}, we numerically fit the lensed dephasing using a nonlinear model fit in \textsc{Mathematica}~\cite{Mathematica} based on~\cref{eq:toy_model_phase}.
We evaluate the match of each fitted model to the injected millilensed phase to minimize the difference between the two and obtain the most accurate fit parameters.
The corresponding values for $(A,B)$ are listed in Table~\ref{tab:toy_model_fits}.
In~\cref{fig:toy_model_large_masses_fit}, we compare the lensed dephasing in the frequency domain with the fitted toy model dephasing.
Although the fitted model is an approximation that could be refined with more complex mathematical expressions, it sufficiently captures the oscillatory behavior and the relative changes in phase modulations across the three different lensing cases.

Let us now use the LSA to analytically estimate the systematic error~\cite{Cutler_2007} due to neglecting the effect of millilensing.
At the injected (true) parameters $\bftr{\lambda}$, we define the deviations as $\Delta \Psi(\bftr{\lambda};f) \equiv \Psi_\mathrm{milli}(\bftr{\lambda};f) - \Psi_\mathrm{ppE}(\bftr{\lambda};f)$, resulting in $\Delta \Psi(\bftr{\lambda};f) = A \sin(B f)$, given $\beta_{\rm tr} = 0$ at the injection.
Using Eq.(\ref{eq:sys_error}), we compute the systematic errors in the recovered chirp mass and ppE phase deviation parameter, assuming the recovery phase is modeled by~\cref{eq:toy_model_phase}.
The resultant systematic errors are given by: 
\begin{subequations}
\begin{align}
    \Delta \mathcal{M} =&-\frac{128}{5}\frac{A u_{0,\mathrm{tr}}^5 \mathcal{M}_\mathrm{tr}}{{\det \hat{\Gamma}(b)}} I_{\mathcal{M}}(b,Bf_0),  \label{eqn:sysbias_mchirp} \\
    \Delta \beta = &\frac{{A u_{0,\mathrm{tr}}^{-b}}}{{\det \hat{\Gamma}(b)}} I_{\beta}(b,Bf_0)\,,
    \label{eqn:sysbias_beta}
\end{align}    
\end{subequations}
where $u_{0,\mathrm{tr}} = (\pi\mathcal{M}_\text{tr} f_0)^{1/3}$, with $\mathcal{M}_\text{tr}$ the injected chirp mass and $f_0$ a characteristic frequency set to 20 Hz. 
We have defined $\det \hat{\Gamma}(b)= I(17) I(7 - 2b) - I^2(12 - b)$, which represents the determinant of the Fisher matrix for the ppE model, where the parameter $b\in \{-7,-3,-1\}$ corresponds to the $\{-1, 1, 2\}$ PN orders in the recovered ppE model. 
The integrals $I(p)$ and $I_\text{S}(p,q)$ are defined as follows:
\begin{subequations}
\begin{align}
I(p) &=4 \int_{x_{\min }}^{x_{\max }} \frac{d x}{S_n\left(f_0 x\right)} x^{-p / 3}, \label{eqn:moment} \\ 
I_\text{S}(p, q) & =4 \int_{x_{\min }}^{x_{\max }} \frac{d x}{S_n\left(f_0 x\right)} x^{-p / 3} \sin(q x), \label{eqn:sine-moment}
\end{align}    
\end{subequations}
where $S_n\left(f\right)$ is the spectral noise density, $x=f/f_0$ is the frequency normalized with a characteristic frequency $f_0$ (set to 20 Hz here), and $p$ is an integer.
We further introduced the following shorthand notation: 
\begin{subequations}
\begin{align}
    I_{\mathcal{M}}(b,Bf_0) &= \left[ I_\text{S}(12, B f_0) I(7 - 2b) \right. \nonumber \\
        &\quad - I_\text{S}(7 - b, B f_0) I(12 - b) \left.\right], \\
    I_{\beta}(b,Bf_0) &= \left[ I(17) I_\text{S}(7 - b, B f_0) \right. \nonumber\\
        & \quad - I(12 - b) I_\text{S}(12, B f_0)\left.\right].
    \label{eq:moment_integrals_mc_beta}
\end{align}    
\end{subequations}

As expected from the LSA, the systematic errors given by~\cref{eqn:sysbias_mchirp,eqn:sysbias_beta} are linear in $A$, the amplitude of the millilensed dephasing.
When the millilensing period is much shorter than the GW period (implying $Bf_0 \ll 1$), $I_S(p,Bf_0)$ can be expanded, and to linear order in $Bf_0$, one recovers the expected result of the toy model presented in~\cite{rohit:syspaper1}.
In Eq.(\ref{eqn:sysbias_mchirp}), the first term in $I_{\mathcal{M}}(b,Bf_0)$ originates from the variance on the chirp mass $C^{\mathcal{M} \mathcal{M}}$, while the second term arises from the co-variance between the chirp mass and the ppE parameter $C^{\mathcal{M} \beta}$.
Similarly, in Eq.(\ref{eqn:sysbias_beta}), the first term in $I_{\beta}(b,Bf_0)$ is derived from the variance $C^{\beta \beta}$, while the second term arises from the covariance $C^{\mathcal{M} \beta}$.
Expressed another way, the systematic error is influenced by the correlations between the waveform parameters, which change as the ppE index $b$ is varied.
Such correlations are crucial in estimating systematic biases in ppE tests, as emphasized in~\cite{Vallisneri:2012qq,Kejriwal:2023djc,rohit:syspaper1}.
Therefore, the systematic error in ppE parameters varies with the lensing configuration, the ppE index, and the total mass of the binary. 

Evaluating~\cref{eqn:sysbias_mchirp,eqn:sysbias_beta} for the configurations in~\cref{tab:toy_model_fits_large_masses}, we observe the following.
For a given ppE deviation, as the lensing strength increases, so does the systematic error, consistent with the result observed in Sec.~\ref{sec:results - systematic bias}.
Meanwhile, for a given lensing strength, the absolute systematic error increases with the PN order of the ppE deviation, which is also reflected in~\cref{tab:sampling_results_small_masses,tab:sampling_results_large_masses} of Sec.~\ref{sec:results - systematic bias}.
Further, the sign of the systematic error changes as one goes from negative to positive PN order in the ppE deviation.
This change of sign is due to the fact that there is a zero crossing in $\Delta \beta$ for a 0PN ppE deviation, which is easily seen from~\cref{eqn:sysbias_beta}.
The 0PN ppE case is not considered here due to the statistical error also vanishing because of the over-simplicity of our toy model.
The sign of $\Delta \beta$ can be better understood by looking at each term in~\cref{eqn:sysbias_beta} and how each of them contributes.
In~\cref{fig:toy_model_PN_order_dependence}, we show that for $\Delta \beta$, the $C^{\beta \beta}$ term decreases with PN order, while the $C^{\mathcal{M} \beta}$ term increases with PN order, with the terms intersecting at 0PN order.
In other words, for larger PN order, the cross-correlation term contributes more to the systematic error in the ppE deviation.
These analytical results are consistent with what we observed from the nested sampling analysis in Sec.~\ref{sec:results - systematic bias}.

To characterize the statistical significance of the systematic bias, as we did in Sec.~\ref{sec:results - systematic bias}, we compute the Bayes factor between the ppE and GR models, as well as the fitting factor of the ppE model.
To obtain an analytic expression for the Bayes factor, we use the Laplace approximation wherein the posterior is well represented by a Gaussian near the maximum posterior point~\cite{gregory_2005}.
The Laplace approximation is typically valid for large SNRs, when the prior becomes insignificant and when the likelihood can be approximated with the Fisher approximation.
We use the Laplace approximation for the Bayes factor, given by~\cref{eqn:laplace}, and estimate $\mathrm{BF}^\text{ppE}_\text{GR}$ as~\cite{rohit:syspaper1}
\begin{subequations}
\begin{align}
    \text{BF}_{\rm ppE,GR} &\approx \dfrac{\sqrt{2 \pi } \sigma_{\beta}}{\beta_{\max}-\beta_{\min}} \exp \left[ \dfrac{1}{2} \left( \dfrac{\Delta \beta}{\sigma_{\beta}} \right)^2 \right], \label{eqn:BF_toy} \\
    \left( \dfrac{\Delta \beta}{\sigma_{\beta}} \right)^2 &= \dfrac{\rho^2 A^2 I^2_{\beta}(b, B f_0)}{I(7)I(17)\det \hat{\Gamma}(b)},
\end{align}      
\end{subequations} 
where $(\beta_{\min},\beta_{\max})$ correspond to the minimum and maximum values of the uniform prior for $\beta_{\rm nPN}$, $\Delta \beta$ and $\sigma_{\beta}$ represent the systematic and statistical errors of $\beta_{\rm nPN}$, respectively, and $\rho$ denotes the SNR of the signal.
The 1-sigma statistical error $\sigma_{\beta}$ is given by (see~\cite{rohit:syspaper1})
\begin{equation}
    \sigma_{\beta} = \frac{u_{0, \rm{tr}}^{-b}}{\rho}\sqrt{\frac{I(7)I(17)}{\det \hat{\Gamma}(b)}}.
\end{equation}

The Bayes factor $\mathrm{BF}^\text{ppE}_\text{GR}$ has two important contributions: one from the maximum likelihood and another from the Occam penalty.
The maximum likelihood contribution is through the exponential, whose argument grows quadratically with the amplitude of the millilensed dephasing, as well as with the SNR.
Further, given that both $\Delta \beta$ and $\sigma_{\beta}$ vary with the PN order of the ppE term, so too does the maximum likelihood contribution.
The Occam penalty contribution essentially depends on the ratio of the statistical error to the prior volume of the ppE parameter.
Due to the dependence on the statistical error, the Occam penalty scales inversely with the SNR.
The Occam penalty also changes with the PN order of the ppE deviation, since both the statistical error and the prior volume change with the PN order.

We use~\cref{eqn:fitting_factor} to compute the fitting factor of the ppE model $\mathrm{FF}_{\rm ppE}$ in the LSA, and obtain
\begin{align}
\begin{split}
    \mathrm{FF}_{\rm ppE} 
    &= \mathcal{F} + \dfrac{A^2}{2 (\det \hat{\Gamma}(b))^2 I(7)} \left[I^2_{\mathcal{M}}(b,Bf_0) I(17) \right. \\
    &\quad +I^2_{\beta}(b,Bf_0) I(7-2b) \\
    &\quad + 2 I_{\mathcal{M}}(b,Bf_0) I_{\beta}(b,Bf_0) I(12-b) \left. \right].    
    \label{eqn:FF_toy}
\end{split}
\end{align}    
Recall that the first term in~\cref{eqn:FF_toy} is the match between the ppE model and the signal at the injected parameters, given by~\cref{eqn:match}. 
When the dephasing induced by millilensing is small, expanding in $\Delta \Psi \ll 1$ to the lowest order, the mismatch is given by~\cite{rohit:syspaper1}
\begin{subequations}
\begin{align}
    1-\mathcal{F} 
    &= \dfrac{A^2}{4} \left[ 1- \dfrac{I_C(7,2Bf_0)}{I(7)} \right]. \label{eqn:match-toy} 
\end{align}    
\end{subequations}
In~\cref{eqn:match-toy}, we have also defined the cosine weighted moment (similar to~\cref{eqn:sine-moment}), given by
\begin{align}
I_\text{C}(p, q) =4 \int_{x_{\min }}^{x_{\max }} \frac{d x}{S_n\left(f_0 x\right)} x^{-p / 3} \cos(q x). \label{eqn:cosine-moment}
\end{align}
As expected, the fitting factor and match are invariant with a rescaling of the SNR, and when the amplitude of the millilensed dephasing vanishes, they become unity.
Furthermore, given that the second term in~\cref{eqn:FF_toy} is positive semi-definite (manifestly seen from~\cref{eqn:fitting_factor}), we have that $\mathrm{FF}_{\rm ppE} \geq \mathcal{F}$.
Put another way, when the systematic errors from neglecting millilensing effects are non-zero, the ppE model evaluated at the maximum likelihood parameters matches the signal better than at the injected parameters.

To demonstrate the dependence of the systematic bias on the ppE index and the lensing configuration (inferred from Sec.~\ref{sec:results - systematic bias}), we evaluate~\cref{eqn:BF_toy,eqn:FF_toy} for the $(24,16)\msun$ system at an SNR of 30 and use the fitted values of $(A,B)$ presented in~\cref{tab:toy_model_fits_large_masses}.
We compute the systematic error $\Delta\beta$, the normalized systematic error $\Delta\beta/\sigma_\beta$, the fitting factor between the recovery ppE model $\mathrm{FF_{ppE}}$, and the injected millilensed model, and the Bayes factor for model selection between ppE and recovery GR models $\text{BF}^{\rm ppE}_{\rm GR}$. 
Similar to our nested sampling analysis, we find that systematic error $\Delta\beta$ increases with lensing strength at a given PN order, and the sign of the systematic error flips when moving from -1PN to positive PN orders.
Additionally, systematic error increases with PN order, for a fixed lensing strength. 
We also observe that fitting factor $\mathrm{FF_{ppE}}$ decreases with lensing strength, while the Bayes factor increases with lensing strength (see~\cref{tab:sampling_results_small_masses,tab:sampling_results_large_masses}). 
These trends align with our findings from the PE study, demonstrating the feasibility of the toy model despite its simplicity.

Several limitations, however, should be noted. 
At a given lensing strength, the nested sampling analysis results in a lower $\mathrm{FF_{ppE}}$ compared to the toy model.
This discrepancy likely arises from different assumptions in the toy model calculations. 
The toy model relies on only two parameters and evaluates results for a single LIGO detector, whereas the nested sampling results are obtained from inference on the complete set of 15 BBH parameters, with an extra ppE parameter, and using a network of three LIGO-Virgo detectors.
Additionally, the toy model employs the LSA, which may be inadequate for the millilensed injections considered in our study, as the amplitude of the millilensed dephasing is not small. 
For the small lensing configuration, the absolute amplitude is $\sim 0.43$ radians, while for the large lensing configuration, it is $\lesssim 0.84$ radians.
Since the systematic error is obtained through an expansion in $\Delta \Psi \ll 1$ and a linearization of the waveforms about the maximum likelihood\footnote{One first linearizes the waveform $h(\bfml{\lambda})$ about $\bfml{\lambda}$ using ``$\Delta \BF{\lambda} \ll 1$'', which strictly means $\vert \partial_{\lambda^i} \partial_{\lambda^j} h (\bftr{\lambda}) (\Delta \lambda^i) (\Delta \lambda^j) \vert \ll \vert \partial_{\lambda^i}h(\bftr{\lambda}) \Delta \lambda^i \vert$. Then, one further linearizes the difference between the signal and the waveform for small dephasing. Since $\Delta \lambda^i$ is linear in the dephasing (see discussion in~\cite{rohit:syspaper1}), the perturbative regime of the bivariate expansion is simply controlled by $\Delta \Psi \ll 1$. } (i.e., ``$\Delta \BF{\lambda} \ll 1$''), neglecting higher order contributions to the systematic error results in an underestimate of $\Delta \beta$.

From the BF results, we observe a clear relation between the value of the BF and the PN order.
This is tied to how well $\beta_{\rm nPN}$ is constrained at a given PN order, as well as the $\beta_{\rm nPN}$ prior range which is larger at higher PN orders.
A slight increase in the BF with lensing is observed for each PN order; however, this increase is not as significant as what we observed in the nested sampling results (see~\cref{tab:sampling_results_small_masses,tab:sampling_results_large_masses}).
Overall, the toy model has effectively clarified the behavior associated with PN orders, particularly the sign change in the systematic errors, and has confirmed the trends observed in the nested sampling results.
Despite its simplicity and with potential for further enhancement, the toy model has successfully facilitated a direct comparison between the two models (ppE and millilensing), which otherwise lack a straightforward one-to-one mapping.

\begin{table*}
\centering 
\begin{ruledtabular}
\begin{tabular}{lrrrrrrr}
PN & Lensing &  $\Delta\beta$ & $\vert\Delta \beta/\sigma_\beta\vert$ & $\text{FF}_\text{ppE}$ &  $\text{BF}^\text{ppE}_\text{GR}$  \\
\colrule
-1  & Small  & $6.716 \times 10^{-9}$  &  0.71 &  0.954 &  $2.245\times 10^{-1}$ \\
-1  & Medium & $6.239 \times 10^{-7}$  & 1.45 & 0.914  &  $2.264 \times 10^{-1}$ \\
-1  & Large  & $4.278 \times 10^{-6}$  & 1.695  & 0.827 &  $3.084\times10^{-1}$ \\
\colrule
1  & Small  & $-1.814 \times 10^{-5}$ & 0.44 &  0.954  & $2.003\times 10^{-3}$ \\
1  & Medium & $-2.501 \times 10^{-4}$ & 0.87 & 0.914 & $2.020 \times 10^{-3}$ \\
1  & Large  & $-1.581 \times 10^{-3}$ & 0.727 & 0.827  &  $2.752\times10^{-3}$ \\
\colrule
2  & Small  & $-1.339 \times 10^{-4}$ & 0.32 & 0.954 & $3.340\times 10^{-4}$ \\
2  & Medium & $-1.942 \times 10^{-3}$ & 0.59 & 0.914 & $3.368 \times 10^{-4}$ \\
2  & Large  & $-1.195\times 10^{-2}$ & 0.281 & 0.827 & $4.589\times10^{-4}$ \\
\end{tabular}
\end{ruledtabular}
\caption{\justifying
Systematic error estimates, along with the ppE fitting factor, and Bayes factor in favor of ppE over GR (see~\cref{eqn:sysbias_beta,eqn:sysbias_mchirp,eqn:BF_toy,eqn:FF_toy} obtained using the toy model. 
We present the values of the above quantities for the large mass system $(24,16)\msun$, and for -1PN, 1PN, and 2PN ppE recoveries, given small, medium, and large lensing injections.
}
\label{tab:toy_model_fits_large_masses}
\end{table*}


\section{\label{sec:conclusion}Conclusion}
In this work, we have studied the systematic biases in ppE tests of GR induced by neglecting gravitational millilensing effects in GW signals detectable by ground-based detectors. 
We specifically studied biases induced by three lensing configurations, characterized by the relative misalignment of the GW source: small-, medium- and large-lensing, corresponding to relative source positions $y\in(0.9, 0.6, 0.3)$, respectively.
We considered two sets of BBH source masses, $M_{\rm{tot}}=\{20, 40\} M_\odot$, and we varied the SNR of the signals in the range $\textrm{SNR}\in(10, 30, 60)$, to probe how the systematic biases depend on the signal duration and SNR.
We performed a Bayesian inference incorporating nested sampling analysis, using ppE recovery models with phase deviations at the -1PN, 1PN, and 2PN orders, while assuming lensing effects to be absent in the recovery model.

Our results indicate that neglecting millilensing effects in the large-lensing case ($y=0.3$) leads to observable systematic biases for signals with SNRs of 30 (or larger) with Bayes factors showing a weak preference for the ppE model over the GR recovery model $10>\text{BF}^\text{ppE}_\text{GR}>1$ and
significant SNR loss quantified by the fitting factor $FF_{\textrm{ppE}}\simeq 0.81$.
The systematic bias was even more important at the higher SNR of 60, where statistical errors are much smaller as compared to systematic errors.
When the SNR is 60, there is an overwhelming preference for the ppE model over the GR model.
Given that there is still a significant loss of SNR for a louder signal, we find that the biases are cases of a Weak Inference of No GR Deviation II, passing the model selection test, but failing an SNR residual test (due to the loss in fitting factor).

In all cases studied, we did not find any instance of an Incorrect Inference of a GR Deviation (when both model selection and SNR residual tests are passed), owing to the significant SNR loss that would lead to a failure of the SNR residual test. Thus, the biases point to the fact that the ppE model is not an accurate fit to the signal data, and instead stems from mismodeling the gravitational waveform, which is indeed the case.
Our findings suggest that neglecting millilensing effects in GWs observable by LIGO-Virgo detectors will not be misinterpreted as false positive deviations from GR if one uses a ppE test and a residual SNR test \textit{simultaneously}.
Notably, our simple toy model for a millilensed GW signal provides analytic insights into the properties of the systematic error observed in our nested sampling analysis, allowing for a direct comparison of the ppE and millilensing models which lack a straightforward one-to-one mapping.
We found that the systematic biases are dependent on the PN order of deviations considered, with positive (negative) biases occurring at negative (positive) PN orders.

Our study has a few limitations but there are several ways to overcome them in future work. 
First, our study is limited by the geometric optics approximation used to model GW lensing, which limits the lower mass of the lenses to $\mathcal{O}(10^3)M_\odot$.
A natural way to overcome this limitation is to analyze the systematic biases arising from GW lensing in the wave optics regime (microlensing), which could account for a population of lenses in a lower mass range. 
Furthermore, the signals analyzed in this study are limited to transient GWs detectable with current ground-based GW detectors at O4 sensitivity. 
Future work could assess the systematic biases arising from neglecting millilensing in next-generation detectors, which are expected to detect signals with much higher SNRs, and for which systematic biases will become even more important. 

\acknowledgements
A.~L. and O.~A.~H. acknowledge support by grants from the Research Grants Council of Hong Kong (Project No. CUHK 14304622 and 14307923), the start-up grant from the Chinese University of Hong Kong, and the Direct Grant for Research from the Research Committee of The Chinese University of Hong Kong. 
R.~S.~C.~and N.~Y.~acknowledge support from the Simons Foundation through Award No. 896696, the National Science Foundation (NSF) Grant No. PHY-2207650 and NASA through Grant No. 80NSSC22K0806. 
The authors are grateful for computational resources provided by the CIT and LHO clusters of the LIGO Laboratory and supported by National Science Foundation Grants PHY-0757058 and PHY-0823459. This material is based upon work supported by NSF's LIGO Laboratory which is a major facility fully funded by the National Science Foundation. This manuscript has LIGO DCC number LIGO-P2400481.
We thank Yiqi Xie for the discussion on the ppE implementation and we are grateful to Emanuele Berti, Justin Janquart, Samson Leong, and Tomek Baka for their valuable feedback and discussions.


\appendix

\section{Supplementary Material For Parameter Estimation Resuls\label{appendix_PE}}
In this appendix, we present supplementary material related to the Bayesian inference results discussed in~\cref{sec:results - systematic bias}.
We compare the significance of deviations from GR in GW amplitude and phase, presenting that the primary contribution to deviations comes from changes in the GW phase.
Additionally, we provide a comparison of systematic bias results estimated from the maximum posterior and median values, alongside the maximum likelihood point presented in~\cref{tab:sampling_results_small_masses,tab:sampling_results_large_masses}.
Finally, we show results of the chirp mass and ppE parameter $\beta_{\rm nPN}$ obtained for the large BBH mass system, analogous to the results from the small-mass system shown in~\cref{fig:mchirp_beta_vary_lens_combined}.

\begin{figure}
    \centering
    \includegraphics[width=0.85\linewidth]{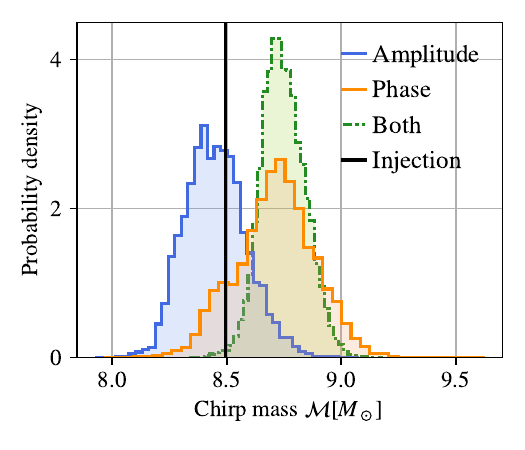}
    \caption{\justifying Comparison of amplitude and phase modulations.
    The results were obtained from three signals modeled as (i) lensed amplitude with unlensed phase (blue), (ii) unlensed amplitude with lensed phase (orange), and (iii) both amplitude and phase lensed (green).
    We observe that case (i) aligns closely with the injected value (vertical black line).
    Notably, phase modulation is the primary contributor to the observed bias.}
    \label{fig:AM_PM}
\end{figure}

As discussed in~\cref{subsec:review-bias}, deviations from GR could change the GW amplitude and phase, with the ppE framework accounting for both types of deviations.
Throughout this work, we focus exclusively on potential deviations in the GW phase, which we demonstrate to be the dominant form of deviation.
Consequently, if any deviations are detected, it is likely that phase changes will be more pronounced than amplitude changes.
This is illustrated in~\cref{fig:AM_PM}, where we performed injections of signals modeled as follows: (i) lensed GW amplitude with unlensed GW phase, (ii) unlensed GW amplitude with lensed GW phase, (iii) both amplitude and phase being lensed. 
The results show that the posterior distribution of the signal with only lensed amplitude (i) does not exhibit notable bias away from the injection value, whereas the phase-only deviations show substantial bias, consistent with results from the signal where both amplitude and phase were lensed.
These findings highlight the importance of prioritizing phase deviations in our analysis.

\begin{figure*}[ht!]
    \centering
    \includegraphics[width=0.8\linewidth]{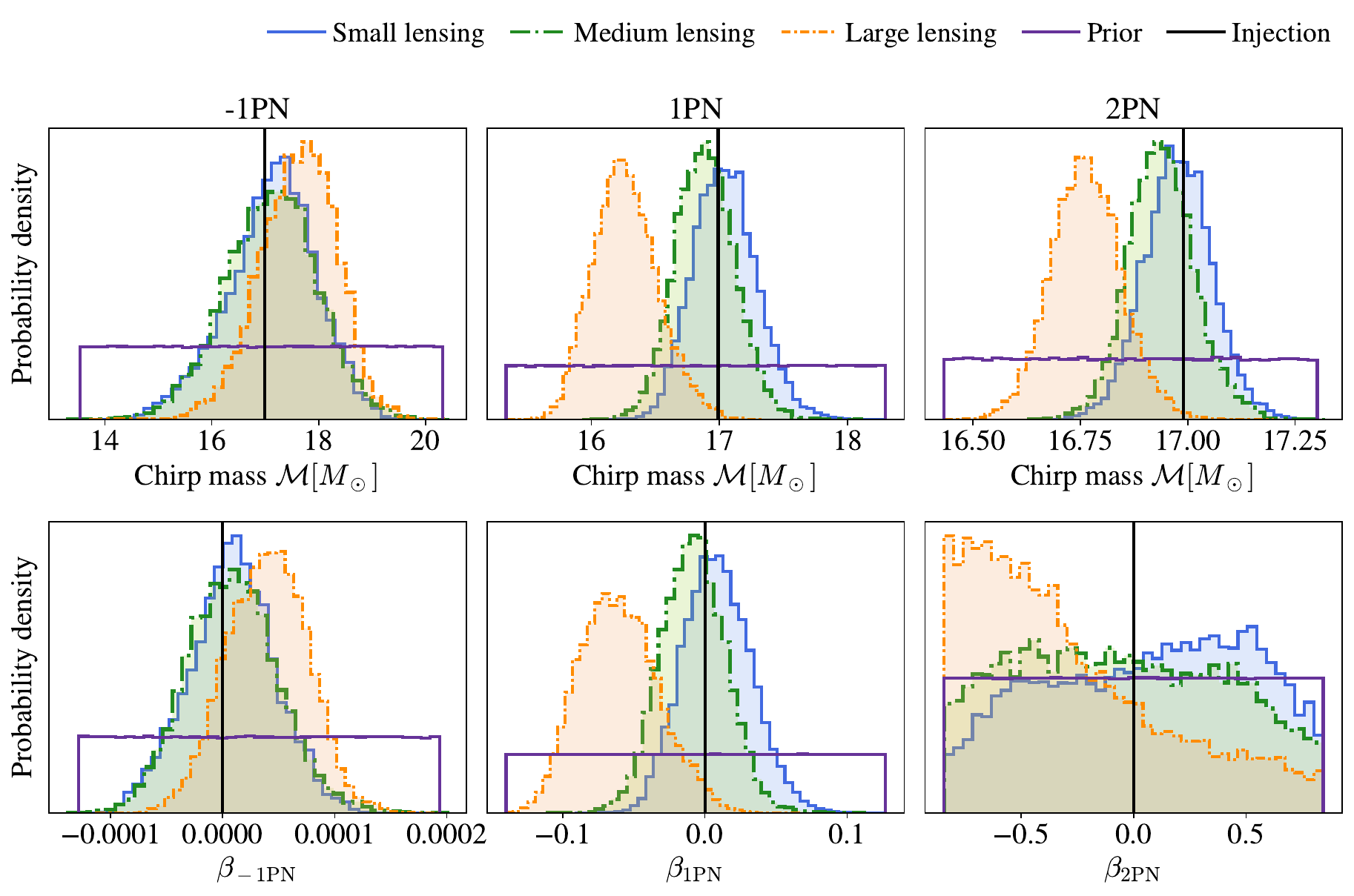}
    \caption{\justifying Comparison of chirp mass $\mathcal{M}$ and $\beta_{\rm nPN}$ recovery for a large-mass BBH system ($24, 16$) $M_\odot$ with an injected SNR of 30 and varying lensing strength.
    The small-, medium- and large-lensing cases correspond to $y\in(0.9, 0.6, 0.3)$, respectively. 
    Similar to the small-mass BBH results, the large-lensing case (orange) shows a bias away from GR, with a preference for larger $\mathcal{M}$ and $\beta_{\rm nPN}$ values than the injection (marked by the vertical black line) at -1PN order, and smaller values at positive PN orders. 
    The small- and medium-lensing cases do not exhibit significant bias.}
    \label{fig:large_mass_BBH_mchirp_beta_vary_lens}
\end{figure*}

Furthermore, we present the PE results for the large mass $(24, 16)M_\odot$ system, complementing the results for the small mass system discussed in~\cref{sec:results - systematic bias}.
We show a comparison of the chirp mass $\mathcal{M}$ and $\beta_{\rm nPN}$ recovery with varying lensing strength and an injected SNR of 30 in ~\cref{fig:large_mass_BBH_mchirp_beta_vary_lens}.
Similar to the smaller mass $(12, 8)M_\odot$ system, we find that the large lensing case ($y=0.3$) induces significant biases. 
However, unlike the small mass system, the most significant bias is observed at the 2PN order.
Additionally, the statistical errors are larger for the larger-mass system, as indicated by the comparable ratios of systematic error to statistical error for both small and large source mass systems, as shown in~\cref{tab:sampling_results_small_masses,tab:sampling_results_large_masses}.

As introduced in~\cref{sec:introduction}, we quantified systematic biases using a likelihood-based measure of the systematic errors. 
We note, however, that there is a numerical uncertainty associated with the maximum likelihood point estimation from Bayesian analysis estimates. 
In~\cref{tab:sys_error_estimates_comparison} we present a comparison of systematic error estimates obtained from three different methods: maximum likelihood point estimates, maximum posterior estimates, and the median of $\beta_{\rm{nPN}}$.

\begin{table*}
\centering 
\begin{ruledtabular}
\begin{tabular}{lrrcrcrcrc}
PN & Lensing &  $\Delta\beta_{\rm{maxL}}$ & $\vert\Delta \beta/\sigma_\beta\vert_{\rm{maxL}}$ & $\Delta\beta_{\rm{maxP}}$ & $
\vert\Delta \beta/\sigma_\beta\vert_{\rm{maxP}}$  & $\Delta\beta_{\rm{Med}}$ & $\vert\Delta \beta/\sigma_\beta\vert_{\rm{Med}}$ \\
\colrule
-1 & S  & $4.033\times 10^{-6}$ & 0.529 & $-2.337\times 10^{-6}$ & 0.307 & $-9.827\times 10^{-7}$ & 0.129 \\
   & M & $3.845\times 10^{-6}$ & 0.392 & $7.657\times 10^{-8}$ & 0.008 & $3.367\times 10^{-6}$ & 0.343 \\
   & L  & $1.495\times 10^{-5}$ & 1.343 & $2.144\times 10^{-5}$ & 1.926 & $1.838\times 10^{-5}$ & 1.651 \\
\colrule 
1 & S   & $6.029 \times 10^{-4}$  & 0.033 & $1.887\times 10^{-3}$ & 0.103 & $-1.482\times 10^{-3}$ & 0.081 \\
  & M  & $-3.141 \times 10^{-3}$ & 0.050 & $ 4.485\times 10^{-4}$ & 0.019 & $-5.407\times 10^{-3}$ & 0.226 \\
  & L  & $-2.273 \times 10^{-2}$ & 0.969 & $-2.536\times 10^{-2}$ & 1.082 & $-2.588\times 10^{-2}$ & 1.104  \\
\colrule
2 & S   & $-1.542 \times 10^{-1}$ & 0.279 & $-1.070\times 10^{-1}$ & 0.154 & $-6.594\times 10^{-3}$ & 0.009  \\
  & M  & $2.583 \times 10^{-1}$ & 0.192 & $3.066\times 10^{-1}$ &  0.454 & $5.819\times 10^{-2}$ & 0.086 \\
  & L  & $-5.083 \times 10^{-1}$ & 1.653 & $-6.636\times 10^{-1}$ & 1.020 & $-3.685\times 10^{-1}$ & 0.566 \\
\end{tabular}
\end{ruledtabular}
\caption{\justifying
Comparison of the systematic errors and systematic over statistical error estimates from the maximum likelihood point $\Delta\beta_{\rm{maxL}}$, the maximum posterior point $\vert\Delta \beta/\sigma_\beta\vert_{\rm{maxP}}$, and the median point $\Delta\beta_{\rm{Med}}$. 
For the small mass system $(12,8)\msun$, we present the values for -1PN, 1PN, and 2PN ppE recoveries, given small, medium, and large lensing injections.
The FF calculations showed no significant change across the three methods (with differences less than 0.01).
}
\label{tab:sys_error_estimates_comparison}
\end{table*}
\begin{figure}[ht!]
    \centering
    \includegraphics[width=0.8\linewidth]{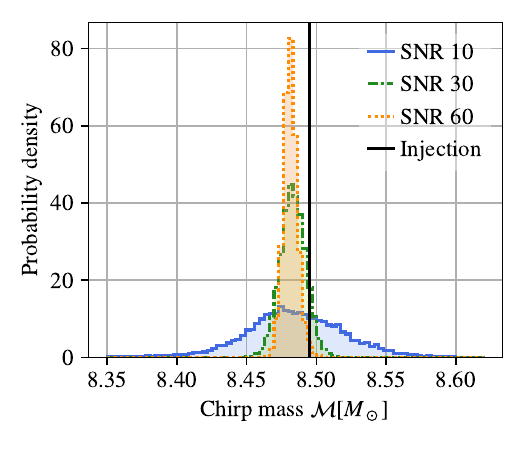}
    \caption{\justifying 2PN chirp mass recovery -- comparison for different injected SNRs.
    As the SNR increases, the posterior width becomes narrower (statistical error decreases).
    The injection value (vertical black line) falls outside of the 90\% CI of the posterior obtained from the signal with an SNR of 60. }
    \label{fig:mchirp_snr_comparison_2PN}
\end{figure}
Lastly, we present complementary results of the biases observed in signals at SNRs of 10, 30, and 60.
We present the recovered chirp mass posterior distribution at the three different SNR values in~\cref{fig:mchirp_snr_comparison_2PN}.
All results were obtained from a large-lensing ($y=0.3$) injection. 
As the SNR increases, the statistical error decreases, consequently the systematic biases become more prominent.
At the SNR of 60, the injected true value falls outside the 90\% credible interval, highlighting the importance of accounting for biases in signals with high SNR.

\begin{figure*}[ht!]
    \centering
    \includegraphics[width=\linewidth]{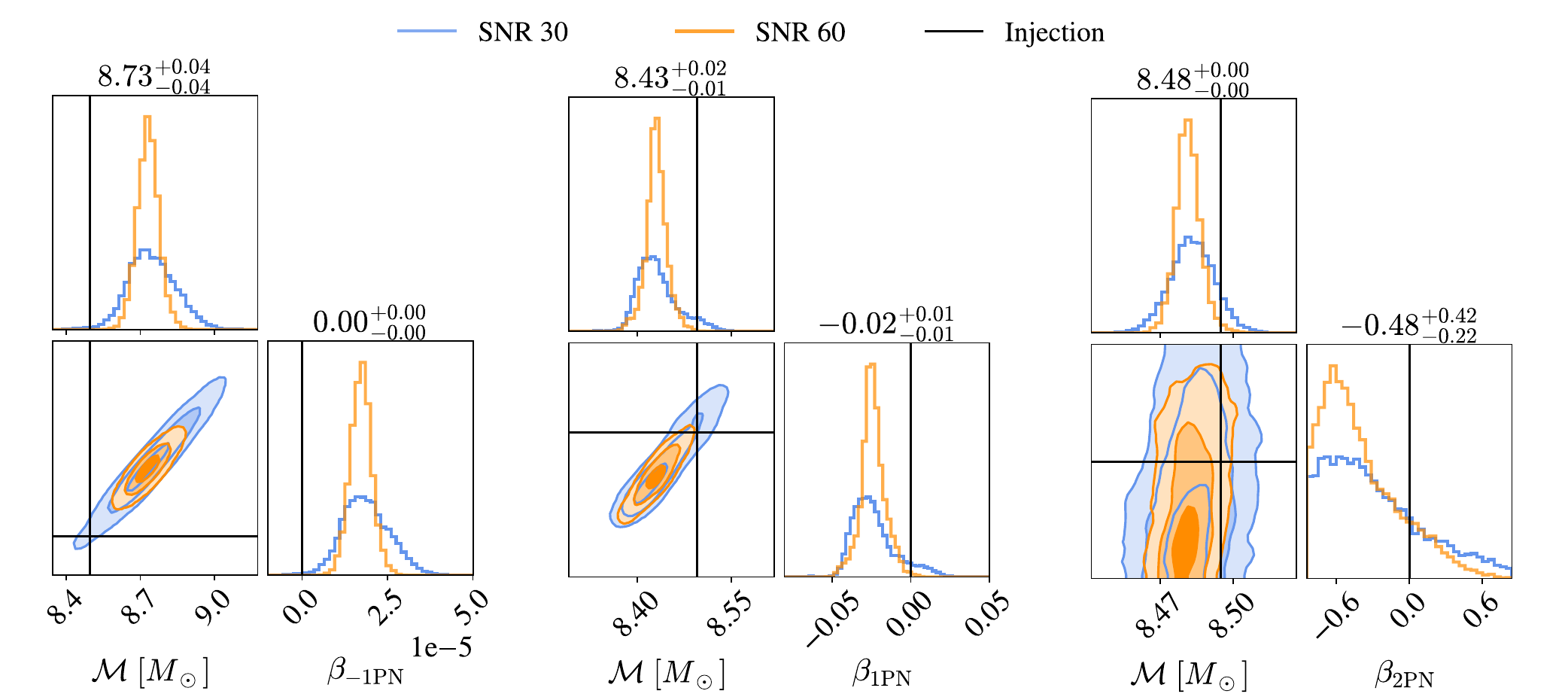}
    \caption{\justifying Recovery of chirp mass $\mathcal{M}$ and $\beta_{\rm nPN}$ at -1PN, 1PN and 2PN order for small mass $(12, 8)M_\odot$ system in the large lensing $y=0.3$ case and varying injected $\mathrm{SNR}\in(30, 60)$. 
    As the SNR increases, the statistical error decreases, hence the systematic bias is more prominent for higher SNR signals.
    It can also be noted that the correlation between $\mathcal{M}$ and $\beta_{\rm nPN}$ is strongest at -1PN order and weakens for higher PN orders.}
    \label{fig:corner_snr30_Vs_snr60_combined}
\end{figure*}

\section{Supplementary Material For The Toy Model Calculation\label{appendix_toy_model}}
In this appendix, we present supplementary material for the toy model calculation presented in~\cref{sec:toy_model}.
In our analysis, we compute the fit of our model to the data by using nonlinear regression techniques, adjusting the parameter $B$ to minimize the differences between the observed phase differences and the fitted values. 
To evaluate the quality of each fit, we calculated the chi-squared statistic to assess how well the model explains the data, with an ideal value close to one indicating a good fit. 
The values obtained from the fit are summarized in~\cref{tab:toy_model_fits}. 

\begin{table}[ht!]
\begin{ruledtabular}
\begin{tabular}{lrr}
source masses & lensing strength & fit parameters \\
($m_1,m_2$)& $y$ & (A, B)\\
\colrule
(24, 16) & small 0.9 & (-0.431, 0.866) \\ 
(24, 16) & medium 0.6 & (0.587, 0.571)\\
(24, 16) & large 0.3 & (0.836, 0.284) 
\end{tabular}
\caption{\justifying Fitted parameters for toy model GW phase.}
\label{tab:toy_model_fits}
\end{ruledtabular}
\end{table}

\begin{figure}[ht!]
    \centering
    \includegraphics[width=\linewidth]{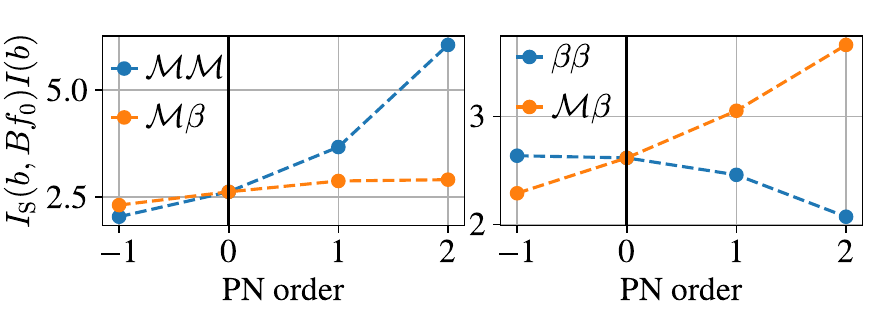}
    \label{fig:toy_model_PN_order_dependence}
    \caption{\justifying Normalized moment integral terms as a function of PN order: $\beta_{\rm{nPN}}$ and $\mathcal{M}$ contributions to the systematic error in chirp mass $\mathcal{M}$ (left) and $\beta_{\rm{nPN}}$ (right). 
    We can note that different contributing terms dominate at -1PN vs 1PN and 2PN orders and they cross at the 0PN order, leading to the change of sign in the systematic error.
    }
\end{figure}

In the analytical derivation of the systematic biases in chirp mass and $\beta$, we observed a sign change when the PN order transitions from negative to positive. 
This phenomenon is directly related to~\cref{eq:moment_integrals_mc_beta} where the integrals derived from the covariance matrix can be expressed as contributions from $C^{\beta \beta}$, $C^{\mathcal{M}\mathcal{M}}$, and $C^{\mathcal{M}\beta}$ terms.  
In particular, we plotted the dependence with the PN order of the individual terms in~\cref{fig:toy_model_PN_order_dependence} which demonstrates the different contributing terms cross at the 0PN order.
Focusing on the systematic error in the chirp mass $\Delta\mathcal{M}$, for $n>0$ PN deviations (where $n=5+b$ with $b$ being the ppE index), the $C^{\mathcal{M} \beta}$ contribution is smaller than the $C^{\mathcal{M} \mathcal{M}}$ contribution.
Meanwhile, looking at $\Delta\mathcal{\beta}$, the $C^{\mathcal{M} \beta}$ contribution is larger than the $C^{\beta \beta}$ contribution for $n>0$ PN deviations.

\bibliography{references}

\end{document}